\documentclass[a4paper,eqsecnum,nofootinbib,showpacs]{revtex4}
\usepackage{slashed,amsmath,amssymb,wrapfig} 
\usepackage{epsfig}
\newcommand{\nn}{\nonumber}
\newcommand{\beq} {\begin{equation}}
\newcommand{\eeq} {\end{equation}}
\newcommand{\beqa} {\begin{eqnarray}}
\newcommand{\eeqa} {\end{eqnarray}}

\newcommand{\as}{{\alpha_s}}

\newcommand{\la}{\Lambda}
\newcommand{\eps}{\epsilon}
\newcommand{\ieps}{i\varepsilon}
\newcommand{\vphi}{\varphi}
\newcommand{\veps}{\varepsilon}
\newcommand{\order}[1]{${\cal O}\left(#1 \right)$}
\newcommand{\morder}[1]{{\cal O}\left(#1 \right)}
\newcommand{\eq}[1]{(\ref{#1})}
\newcommand{\fig}[1]{Fig.~\ref{#1}}
\newcommand{\lsim}{\lesssim}
\newcommand{\gsim}{\gtrsim}

\newcommand{\inv}[1]{\frac{1}{#1}}
\newcommand{\halft}{{\textstyle \frac{1}{2}}}

\newcommand{\sfrac}[2]{{\textstyle\frac{#1}{#2}}}
\newcommand{\ket}[1]{\left\vert{#1}\right\rangle}
\newcommand{\bra}[1]{\langle{#1}\vert}

\newcommand{\com}[2]{\left[{#1},{#2}\right]}
\newcommand{\acom}[2]{\left\{{#1},{#2}\right\}}
\newcommand{\tr}{\mathrm{Tr}\,}

\newcommand{\bs}[1]{\boldsymbol{#1}}
\newcommand{\im}{{\rm Im}}

\newcommand{\T}{{\rm T}}

\newcommand{\Psl}{{\slashed{P}}}
\newcommand{\ksl}{{\slashed{k}}}

\newcommand{\xv}{{\bs{x}}} 
\newcommand{\yv}{{\bs{y}}}

\newcommand{\Pv}{{\bs{P}}}
\newcommand{\gv}{\bs{\gamma}}
\newcommand{\nv}{\bs{\nabla}}

\newcommand{\moe}{{m}}
\newcommand{\Moe}{{M}}
\newcommand{\dm}{{\Delta m^2}}
\newcommand{\kum}{{\,{_1}F_1}}
\newcommand{\xbj}{{x_{Bj}}}

\newcommand{\M}{\mathcal{M}}

\newcommand{\dpi}{\Pi}
\newcommand{\dsi}{\sigma}
\newcommand{\xs}{\sigma_\mathrm{scat}}
\newcommand{\hatxs}{\hat\sigma_\mathrm{scat}}

\begin{document}
{\par\raggedleft \texttt{CCTP-2012-26}\par}
\bigskip{}

\title{Towards a Born term for hadrons}

\author{Dennis D. Dietrich}
\affiliation{Institut f\"ur Theoretische Physik, Goethe-Universit\"at, Max-von-Laue-Str.~1,
D-60438 Frankfurt am Main, Germany}

\author{Paul Hoyer}
\affiliation{Department of Physics and Helsinki Institute of Physics\\ POB 64, FIN-00014 University of Helsinki, Finland}

\author{Matti J\"arvinen}
\affiliation{Crete Center for Theoretical Physics\\ 
Department of Physics, University of Crete\\71003 Heraklion, Greece}

\begin{abstract} 
We study bound states of Abelian gauge theory in $D=1+1$ 
 dimensions using an equal-time, Poincar\'e-covariant framework. The normalization of the linear confining potential is determined by a boundary condition in the solution of Gauss' law for the instantaneous $A^0$ field. As in the case of the Dirac equation, the norm of the relativistic fermion-antifermion ($f\bar f$)
wave functions gives inclusive particle densities. However, while the Dirac spectrum is known to be continuous we find that regular $f\bar f$ solutions exist only for discrete bound-state masses. The $f\bar f$ wave functions are consistent with the parton picture when the kinetic energy of the fermions is large compared to the binding potential. We verify that the electromagnetic form factors of the bound states are gauge invariant and calculate the parton distributions from the transition form factors in the Bjorken limit. For relativistic states we find a large sea contribution at low $\xbj$. Since the potential is independent of the gauge coupling the bound states may serve as ``Born terms'' in a perturbative expansion, in analogy to the usual plane wave {\em in} and {\em out} states.
\end{abstract}


\maketitle

%%%%%%%%%%%%%%%%%%%%%%
\section{Introduction}
%%%%%%%%%%%%%%%%%%%%%%

The hadron spectrum is simpler than one would expect. Deep inelastic scattering shows important contributions from sea quarks and gluons, yet $q\bar q$ mesons and $qqq$ baryons are successfully classified \cite{Beringer:1900zz} in terms of only their valence quark degrees of freedom. Dynamical features such as masses and magnetic moments are
consistent with the nonrelativistic quark model \cite{Richard:2012xw}, even though (light) quarks are known to be ultrarelativistic. Models which take into account relativistic effects have been constructed and successfully compared with data \cite{Basdevant:1984rk,Godfrey:1985xj,Koll:2000ke,Branz:2009cd,Day:2012yh}. Approaches based on relativistic Dyson-Schwinger equations capture many features of hadrons \cite{Roberts:2007ji,Alkofer:2009dm}.

It is well established -- and confirmed by numerical lattice calculations \cite{Colangelo:2010et,Alexandrou:2012hi} -- that hadrons are bound states of Quantum Chromodynamics (QCD). The relative simplicity of the hadron spectrum and the success of quark models motivates us to ask: Is there a systematic approximation scheme of QCD which, at lowest (``Born term'') order, has quark model features? 
Here our ambition is to refrain from either introducing effective quantities (e.g., local fields for hadrons) or postulate potentials beyond the gauge fields of QCD. It may seem that under these conditions the answer should be ``no.'' Our present results 
indicate, however, that this answer is not obviously correct.

The similarities between the spectra of hadrons and atoms induce us to take $\as$
at small momentum transfer as our expansion parameter. Several theoretical and phenomenological studies 
\cite{Brodsky:2002nb,Fischer:2006ub,Deur:2008rf,Aguilar:2009nf,Gehrmann:2009eh,Ermolaev:2012xi,Courtoy:2013qca} find that the strong coupling freezes at a moderate value in the infrared. Perturbation theory provides a well constrained framework for addressing the question we raised above. At lowest order in $\as$ our expansion should resemble quark models, which typically use the Cornell potential \cite{Eichten:1978tg}
\beq\label{qmpot}
V_{QM}(\xv)=c\,|\xv|-C_F\frac{\as}{|\xv|} \ .
\eeq
The second term is due to single gluon exchange and thus arises naturally in our perturbative expansion.
We shall not endeavor to {\em derive} the color confining term $c\,|\xv|$ in \eq{qmpot} -- but neither just postulate it. We rather ask if and how this term is {\em compatible} with the QCD equations of motion.

The interaction \eq{qmpot} is instantaneous. Gluons propagate in time, giving rise to intermediate states with one or several gluons. The Coulomb field $A^0$ of gauge theories is an exception. It has no time derivative in the Lagrangian and is thus instantaneous. We can avoid $\ket{q\bar q\,g},\ldots$ Fock states related to the linear potential $c\,|\xv|$ in \eq{qmpot} only if it is due to Coulomb gluons\footnote{The single gluon exchange term in \eq{qmpot} is instantaneous only for nonrelativistic dynamics. Even photon exchange in QED atoms involves higher Fock states, in frames where the atom moves relativistically \cite{Jarvinen:2004pi}.}. The absence of a time derivative on $A^0$ also implies that the field equations of motion (``Gauss' law'') allow us to express $A^0$ in terms of the propagating fields at each instant of time. In QED Gauss' law specifies, for an $\ket{e^-(\xv_1)e^+(\xv_2)}$ state and in $\nv\cdot\bs{A}=0$ gauge,
\beq\label{gausslaw}
-\nv^2 A^0(\xv) = e\left[\delta(\xv-\xv_1)-\delta(\xv-\xv_2)\right],
\eeq
with the standard solution
\beq\label{pertsol}
A^0(\xv) = \frac{e}{4\pi}\left[\inv{|\xv-\xv_1|}-\inv{|\xv-\xv_2|}\right] \ .
\eeq
The interaction potential is then $\halft[eA^0(\xv_1)-eA^0(\xv_2)]=-\alpha/|\xv_1-\xv_2|$, the familiar Coulomb potential.
However, we may add a {\em homogeneous} solution of \eq{gausslaw} to \eq{pertsol} \cite{Hoyer:2009ep},
\beq\label{homsol}
A^0_\Lambda(\xv)=\la^2\, \hat{\bs{\ell}}\cdot \xv \ ; \hspace{1cm} \bs{A}_\la=0 \ .
\eeq
The constant $\Lambda$ corresponds to a nonvanishing boundary condition in 
the solution of Gauss' law,
\beq\label{Fboundcond}
\lim_{|\xv|\to\infty}F_{\mu\nu}F^{\mu\nu}(\xv)=-2\Lambda^4.
\eeq
The unit vector $\hat{\bs{\ell}}$ must be independent of $\xv$ but can otherwise be 
chosen freely.  
Rotational invariance requires $\hat{\bs{\ell}} \parallel \xv_1-\xv_2$. The potential energy is then
\beq\label{linpot3d}
V(\xv_1-\xv_2) = \halft[eA^0_\la(\xv_1)-eA^0_\la(\xv_2)]= \halft e\la^2 |\xv_1-\xv_2|\ . 
\eeq
An analogous, homogeneous solution of Gauss' law exists in QCD \cite{Hoyer:2009ep}. The parameter $\la$ should vanish for QED to describe data, while in QCD $\la \sim \la_{QCD}$ may be related to the coefficient $c$ of the quark model potential \eq{qmpot}. The homogeneous solution \eq{homsol} exists for charges of any momentum, whereas $A^0$ dominates perturbative exchange only in the case of nonrelativistic dynamics. It is clear from \eq{linpot3d} that the potential $V$ is invariant under translations only for neutral (color singlet) states. Poincar\'e invariance thus requires the bound states to be neutral if $\la \neq 0$. 

We define a neutral fermion-antifermion bound state at equal time ($t=0$) and of 4-momentum $P=(E,\Pv)$ by
\beq\label{emustate}
\ket{P}=\int d\xv_{1}d\xv_{2}\,\bar\psi_{1}(t=0,\xv_{1})\exp\big[i\Pv\cdot (\xv_1+\xv_2)/2\big]\Psi(\xv_{1}-\xv_{2})\psi_{2}(t=0,\xv_{2})\ket{0}_R.
\eeq
Here $\psi_{f}$ is a fermion operator of flavor $f$ in Abelian gauge theory (see \cite{Hoyer:2009ep} for the generalization to QCD), and the $c$-number wave function $\Psi$ has $4\times 4$ Dirac components. 
The boundary condition \eq{Fboundcond} separates charged and neutral states by an infinite (field) energy. This is similar to $D=1+1$ dimensions, where the perturbative potential is linear and physical states are neutral \cite{Coleman:1975pw}. 
The subscript $R$ denotes that we are using the ``retarded vacuum,'' which satisfies $\psi_1(x)\ket{0}_R =\psi_2^\dag(x)\ket{0}_R =0$. This 
eliminates pair production from the vacuum, $H\ket{0}_R=0$, allowing us to describe the bound state in terms of a two-particle Fock state only \cite{Hoyer:2009ep}. It was observed previously \cite{Baltz:2001dp} that scattering amplitudes defined using the retarded vacuum give {\em inclusive} cross sections.

Under a space translation $\xv \to \xv+\bs{d}$ the state \eq{emustate} transforms by a phase $\exp(i\Pv\cdot \bs{d})$, as appropriate for a state of total momentum $\Pv$. Stationarity under time translations imposes
\beq\label{hameig2}
H\ket{P} = E\ket{P}\ .
\eeq
At lowest order in the coupling $e$, neglecting all perturbative contributions, the gauge field is given by \eq{homsol}. 
This contribution can be taken into account by adding an \order{e\Lambda^2} instantaneous interaction term 
to the Hamiltonian, $H \to H + H_\la$, which for neutral states is effectively
\cite{Hoyer:2009ep}
\beq
H_{\la}=-\frac{e\Lambda^2}{4}\sum_{f,f'}\int d\xv\, d\yv\, \psi_f^\dag\psi_f(t,\xv)|\xv-\yv|\psi_{f'}^\dag\psi_{f'}(t,\yv) ,
\eeq
and thus is leading compared to the \order{e^2} perturbative interactions. 
For $\ket{P}$ to be an eigenstate of $H$ at \order{e\Lambda^2} the wave function $\Psi(\xv_{1}-\xv_{2})$ should satisfy
the bound-state equation
\beq\label{bse1}
i\nv_x\cdot\acom{\gamma^0\gv}{\Psi(\xv)}-\halft \Pv\cdot\com{\gamma^0\gv}{\Psi(\xv)}+m_1\gamma^0\Psi(\xv)-m_2\Psi(\xv)\gamma^0 = \big[E-V(\xv)\big]\Psi(\xv),
\eeq
with the potential $V(\xv)$ given by \eq{linpot3d}.

The wave functions $\Psi$ and the energy eigenvalues $E$ that solve the bound-state equation \eq{bse1} depend on the 3-momentum $\Pv$. In this equal time, Hamiltonian formalism boost invariance (as well as time translation invariance) is a dynamical symmetry. Because the field theory 
is Poincar\'e invariant and is solved at lowest order in the coupling $e$, with the Poincar\'e invariant boundary condition \eq{Fboundcond}, we expect the state \eq{emustate} to be covariant. In \cite{Dietrich:2012iy} we verified this in $D=1+1$ dimensions using the boost generator $M^{01}$. The boosted state satisfies the bound-state equation \eq{bse1} with the appropriately shifted momentum, i.e.,
\beq\label{boostop}
(1-id\xi M^{01})\ket{P^0,P^1}=\ket{P^0+d\xi P^1,P^1+d\xi P^0}.
\eeq
 Also in $D=3+1$ dimensions the energy eigenvalues of \eq{bse1} have the required dependence on the momentum \cite{Hoyer:1986ei},
\beq\label{edep}
E = \sqrt{M^2+\Pv^2}.
\eeq
This indicates that the states are Poincar\'e covariant also in four-dimensional space-time.

So far we discussed bound-state solutions in the presence of only the nonperturbative field \eq{homsol}, which gave the linear potential \eq{linpot3d}. We conjecture that perturbative corrections can be taken into account in the standard way. Then a formally exact expression for an $S$-matrix element is 
\beq\label{smatrix}
S_{fi}={}_{\rm out}\bra{f}T\big[\exp(-i \int d^4x\,\mathcal{H}_I)\big]\ket{i}_{\rm in} \ .
\eeq
Usually the particles in the $in$ and $out$ states are taken to be free fields. In the present framework they are (collections of) the \order{\alpha^0} bound states \eq{emustate} at asymptotic times $t=\mp \infty$. 
In effect, the $S$ matrix is perturbatively expanded around Born states which are zeroth order approximations of QCD hadrons, reminiscent of quark model states and Poincar\'e covariant.  We plan to study the properties of this perturbative expansion in future work.

The present paper is a sequel to our study \cite{Dietrich:2012iy} of Poincar\'e invariance. Here we examine other features of the solutions to the bound-state equation \eq{bse1} in $D=1+1$. These wave functions have several unusual properties, some of which are shared with the (previously known) solutions of the Dirac equation. The $D=1+1$ wave functions can be expressed in terms of confluent hypergeometric functions, which allows detailed analytic and numerical studies. We expect that the results in this paper shed some light also on the properties of the solutions of \eq{bse1} in $D=3+1$.

\vspace{2mm}

In the next section we study the solutions of the Dirac equation
for a linear potential in $D=1+1$ dimensions. We show how, for nearly nonrelativistic dynamics ($m \gsim Ze$), the solutions  
agree with those of the corresponding Schr\"odinger equation
at small fermion separations $|x|$, 
but reflect pair production at separations where $V(x) \gsim 2m$. 
In Sec.~\ref{ffbarsec} we find the analytic solutions of the $f\bar f$ bound-state equation \eq{bse1} in $D=1+1$ dimensions. Similarly to the Dirac wave functions they are not square integrable since their norm tends to a constant at large  
$|x|$. 
The $f \bar f$ wave functions, however, are generally singular at $V(x)=E \pm P^1$.
Solutions that are regular at these points\footnote{When the fermion masses are unequal, 
$m_1 \neq m_2$ in \eq{bse1}, the wave functions are fully regular only in the infinite momentum frame, $P^1 \to \infty$.} 
exist only for discrete bound-state masses \cite{Geffen:1977bh}. This differs 
qualitatively from the Dirac equation, whose solutions 
are regular at all $x$, 
giving a continuous mass spectrum. In Sec.~\ref{dualsec} we show that the normalization of the wave function at $x=0$ can be determined by requiring duality between the bound state ($\delta$ distribution) and fermion-loop contributions to the imaginary parts of current propagators. A self-consistent normalization is obtained in all Lorentz frames and for all currents.
For highly excited  
bound states the wave functions 
[at low $|x|$, hence small $V(x)$] 
turn into plane waves of positive energy fermions as expected in the parton model. In Sec.~\ref{dissec} we 
evaluate the electromagnetic form factors and show that they are gauge invariant  
(in any dimension). 
In Sec.~\ref{dis} we express deep inelastic scattering in terms of transition form factors in the Bjorken limit.
For relativistic states (small fermion masses) the $D=1+1$ parton distributions 
grow large at small $\xbj$, with $\xbj f(\xbj) \propto \log^2(\xbj)$. 
Our conclusions are given in Sec.~\ref{concsec}. 

%%%%%%%%%%%%%%%%%%%%%%
\section{Dirac equation in $D=1+1$} \label{diracsec}
%%%%%%%%%%%%%%%%%%%%%%

Some of the novel properties of the $f\bar f$ states  
that we study in $D=1+1$ dimensions, notably the asymptotically constant norm
of their wave functions, 
are shared by electrons bound in an external linear $A^0$ 
potential. Soon after Dirac first proposed his wave equation \cite{Dirac:1928hu} it was realized \cite{nikolsky,sauter,plesset} that solutions of the Dirac equation generally cannot be normalized.
This contrasts with the solutions of  
the Schr\"odinger equation, where the requirement of a finite normalization integral leads to the 
quantization of the energy spectrum. Thus in \cite{plesset} it was shown that the solutions of the Dirac equation in $D=1+1$ have a constant norm as $|x| \to \infty$ for all potentials of the form $A^0(x) = x^n$ (or combinations thereof), where $n \neq 0$ is any positive or negative integer. A similar result holds in $D=3+1$ dimensions when $\bs{A}=0$, for central potentials $A^0(r)$ which are polynomials in $r$ or in $1/r$, with the interesting exception of $A^0(r) \propto 1/r$ \cite{plesset}. The Dirac energy spectrum is thus generally continuous, and the completeness relation involves a continuous set of eigenfunctions \cite{titchmarsh}.

The constant norm of the Dirac wave function reflects pair production in a strong potential. The phenomenon is related to the Klein paradox \cite{Klein:1929zz}, which requires a multiparticle framework for its resolution \cite{Hansen:1980nc}. The pairs can be seen to arise from $Z$ diagrams when the scattering of an electron in an external field is time ordered.  
In the case of static $A^0$ potentials the same
bound-state energies are obtained with retarded as with Feynman 
electron propagators \cite{Hoyer:2009ep}. 
Using retarded propagators 
electrons of both positive and negative energies propagate forward in time, and the bound states are described by single particle Dirac wave functions. Due to the retarded boundary conditions the norm of the Dirac wave function is, however, ``inclusive'' in nature, in analogy with the more familiar concept of inclusive scattering cross sections \cite{Baltz:2001dp}.

The analytic solution of the Dirac equation  
\beq\label{diraceq}
\big[-i\nv\cdot\gv+e\gamma^0 A^0(\xv) +m\big]\left[\begin{array}{c}  \vphi(\xv) \\ \chi(\xv) \end{array}\right] = M\gamma^0 \left[\begin{array}{c}  \vphi(\xv) \\ \chi(\xv) \end{array}\right],
\eeq
in $D=1+1$ for a linear potential was first given in \cite{sauter}. We include the derivation below for completeness and to introduce our notation. We also discuss some numerical properties of the solutions, which appear not to be widely known.

\subsection{General solution} \label{diracsec2}
%%%%%%%%%%%%%%%%%%%%%%

We use a standard two-dimensional representation of the Dirac matrices in terms of Pauli matrices,
\beq\label{Dmatrices}
\gamma^0=\sigma_3\,,\hspace{1cm} \gamma^1=i\sigma_2\,,\hspace{1cm} \gamma^0\gamma^1 = \gamma_5 = \sigma_1 .
\eeq
In QED$_2$ the potential generated by a static source of charge $Ze$ is $V(x)=eA^0(x) = \halft Ze^2|x|$. We use units where $Ze^2=1$, hence
\beq\label{linpot}
V(x) = \halft |x|.
\eeq
In the following all dimensionful quantities can be  given their physical dimensions through multiplication by the appropriate power of $Ze^2$.

The Dirac equation \eq{diraceq} implies
\beqa\label{diraccomp}
-i\partial_x\chi&=&(M-V-m)\vphi, \nn\\
-i\partial_x\vphi&=&(M-V+m)\chi.
\eeqa
We choose the phases such that $\vphi(x)$ is real and $\chi(x)$ is imaginary. This means that the solutions are characterized 
by two real parameters, e.g., $\vphi(0)$ and $-i\chi(0)$. By adding or subtracting the equations at $x$ and $-x$ we may impose that 
\beq\label{dirpar}
\vphi(x) = \eta~\vphi(-x),  \hspace{2cm} \chi(x) = -\eta~\chi(-x), 
\eeq
where $\eta=\pm 1$.
This allows us to consider solutions in the region $x \geq 0$ only. Continuity at $x=0$ requires $\chi(0)=0$ for $\eta=+1$, and $\vphi(0)=0$ for $\eta=-1$. The equations \eq{diraccomp} then ensure that $\partial_x \vphi(0)=0$ for $\eta=1$ and $\partial_x\chi(0)=0$ for $\eta=-1$.

The two first-order equations \eq{diraccomp} give rise to a second-order equation for $\vphi(x)$,
\beq\label{phieq}
\partial_x^2\vphi(x)+\frac{\veps(x)}{2(M-V+m)}\partial_x\vphi(x)+\big[(M-V)^2-m^2\big]\vphi(x)=0,
\eeq
where $\veps(x)\equiv x/|x|$ is the sign function. For $x \to \infty$ the term $V^2\vphi(x)$ must be balanced by $\partial_x^2\vphi(x)$, hence 
\beq\label{asphi}
\vphi(x\to\infty) \sim \exp(\pm i x^2/4).
\eeq 
From \eq{diraccomp} it follows that $\chi(x)$ has a similar asymptotic behavior. Since the norms $|\vphi(x)|$ and $|\chi(x)|$ tend to constants for $x\to\infty$ the Dirac wave functions are not normalizable \cite{nikolsky,sauter,plesset}, unlike the solutions of the nonrelativistic Schr\"odinger equation. As we shall see, the solutions have features which support the interpretation that their norm at large $V(x)$ reflects virtual pair
contributions.

The coefficient of $\partial_x\vphi(x)$ in \eq{phieq} is singular at $M-V+m=0$. Assuming $\vphi(x) \sim (M-V+m)^\alpha$ as $M-V+m \to 0$ we find $\alpha = 0$ or 2. Hence  the general solutions $\vphi(x)$ and $\chi(x)$ have no singularities at finite $x$. Since the wave functions are not square integrable there is no restriction on the eigenvalues $M$. In Sec.~\ref{dirnr} we discuss how the discrete eigenvalues required by the Schr\"odinger equation emerge nevertheless in the nonrelativistic domain ($m\gg 1$). In Sec.~\ref{ortho} we show that any two solutions with different eigenvalues $M$ are orthogonal. 

Since only the combination $M-V(x)$ appears in the Dirac equation \eq{diraccomp} it is convenient to replace $x$ by the variable\footnote{We consider solutions only for $x \geq 0$ in the following. The parity condition \eq{dirpar} gives the solutions for $x<0$.}
\beq\label{tdirac}
\dsi = (M-V)^2, \hspace{2cm} \partial_x = \frac{d\dsi}{dx}\partial_\dsi= -(M-V)\partial_\dsi.
\eeq
The Dirac equation then reads\footnote{Here and in the following we use the shorthand notation 
$\vphi(\dsi)\equiv \vphi[\dsi(x)]$,
and similarly for $\chi(x)$.}, in the region where $M-V(x)>0$,
\beqa\label{diract}
i\partial_\dsi \chi(\dsi) = \Big(1-\frac{m}{\sqrt{\dsi}}\Big)\vphi(\dsi),\nn\\
i \partial_\dsi \vphi(\dsi) = \Big(1+\frac{m}{\sqrt{\dsi}}\Big)\chi(\dsi).
\eeqa

We may combine $\vphi$ and $\chi$ into the single complex function
\beq\label{phidef}
\phi(\dsi) \equiv \big[\vphi(\dsi)+\chi(\dsi)\big]e^{i\dsi}  \hspace{.5cm} {\rm and\ conversely
} \hspace{.5cm} \vphi(\dsi) = {\rm Re}\big[\phi(\dsi)e^{-i\dsi}\big], \hspace{1cm} \chi(\dsi) = i\,{\rm Im}\big[\phi(\dsi)e^{-i\dsi}\big].
\eeq
The second-order equation for $\phi(\dsi)$ is then
\beq
2\dsi\,\partial_\dsi^2\phi+(1 -4i\dsi)\partial_\dsi\phi-2m^2\phi=0 ,
\eeq
with the general solution
\beq
\phi(\dsi)=(a_1+ib_1)\,{_1}F_1(-\halft im^2,\halft,2i\dsi) + (a_2+ib_2)\,
\sqrt{\dsi}\;{_1}F_1(\halft-\halft im^2,\sfrac{3}{2},2i\dsi),
\eeq
where ${_1}F_1$ is the confluent hypergeometric function and the $a_i,\,b_i$ are real constants. 
From \eq{phidef} we find
\beqa
\vphi(\dsi\to 0) = a_1+a_2\sqrt{\dsi} + \morder{\dsi},\nn\\
\chi(\dsi\to 0) = ib_1+ib_2\sqrt{\dsi} + \morder{\dsi}.
\eeqa
Matching the terms of \order{1/\sqrt{\dsi}} in \eq{diract} gives the relations
\beq
b_2=2m a_1, \hspace{2cm} a_2=2m b_1.
\eeq
The general solution of the $D=1+1$ Dirac equation \eq{diraccomp} with the linear potential \eq{linpot} of QED$_2$ is thus given by
\beq\label{phisol}
\psi(\dsi) \equiv \vphi(\dsi)+\chi(\dsi) = \left[(a+ib){_1}F_1\Big(-\frac{im^2}{2},\inv{2},2i\dsi\Big)+(b+ia)2m\,\veps(M-V)\sqrt{\dsi}\,{_1}F_1\Big(\frac{1-im^2}{2},\frac{3}{2},2i\dsi\Big)\right]\exp(-i\dsi),
\eeq
where $a$ and $b$ are real parameters and $\dsi=(M-V)^2$. $\vphi$ and $\chi$ are given by the real and imaginary parts of the right-hand side, respectively. The sign function $\veps(M-V)$ ensures that the solution is valid also in the region where $M-V(x)<0$, since $\sqrt{\dsi}$ in \eq{diract} changes sign at $\dsi=0$. This solution agrees with Eq.~(14) of \cite{sauter}.

The solution \eq{phisol} is valid for $x\geq 0$. The wave functions for $x<0$ are given by the symmetry relations \eq{dirpar}. Continuity at $x=0$ requires that the antisymmetric (real or imaginary) part of the wave function vanishes at $x=0$, which also ensures that the $x$ derivative of the symmetric part vanishes. The positions $\dsi=\dsi_0$ where either the real or imaginary part of the right-hand side in \eq{phisol} vanishes determine the mass eigenvalues $M = \sqrt{\dsi_0}$, since continuity allows to set $x=0$ at $\dsi_0$. The eigenvalues $M$ thus depend on the ratio $a/b$ of the parameters.

The asymptotic behavior of the wave function at large $x$,
\beqa\label{psilimit}
\lim_{x\to\infty}\psi(\dsi) &=& \sqrt{\pi}e^{\pi m^2/4}\left[\frac{a+ib}{\Gamma[\halft(1+im^2)]}+
\frac{a-ib}{\Gamma(1+\halft im^2)}\frac{me^{-i\pi/4}}{\sqrt{2}}\right](2\dsi)^{im^2/2}e^{-i\dsi} \nn\\[2mm]
&+& \frac{\sqrt{\pi}}{\sqrt{2\dsi}}e^{\pi (m^2-i)/4}\left[\frac{a+ib}{\Gamma(-\halft im^2)}-\frac{a-ib}{\Gamma[\halft(1-im^2)]}\frac{me^{i\pi/4}}{\sqrt{2}}\right](2\dsi)^{-im^2/2}e^{i\dsi}+\morder{\inv{\dsi}},
\eeqa
oscillates with $\dsi \simeq x^2/4$ in agreement with the general result \eq{asphi}. Hence the norm of the wave function tends to a constant at high $x$. Unlike in the nonrelativistic limit (the Schr\"odinger equation), the parameters of the solution cannot be determined by a normalizability condition.

\subsection{The nonrelativistic limit  \label{dirnr}}
%%%%%%%%%%%%

The fact that the eigenvalues $M$ of the Dirac equation \eq{diraccomp} depend on the ratio $a/b$ of the parameters in \eq{phisol} raises the question about the approach to the nonrelativistic limit. In $D=1+1$ dimensions for a fixed potential this limit is equivalent to taking $m \to \infty$, scaling simultaneously coordinates and momenta appropriately. For a linear potential the scaling of the coordinate and the binding energy $E_b=M-m$ is
\beq\label{nrscaling}
 x \sim E_b \sim m^{-1/3}. 
\eeq
In the  
nonrelativistic limit the Dirac equation reduces to the Schr\"odinger equation
\beq\label{seq}
 -\inv{2\moe}\partial_x^2 \rho(x)+\frac{1}{2} |x| \rho(x) = E_b\rho(x),
\eeq
whose normalizable solutions are given by the Airy function,
\beq\label{seqAi}
\rho(x) = N\,\mathrm{Ai}[\moe^{1/3}(x-2E_b)] \qquad (x>0), 
\eeq
with $\rho(-x)=\pm \rho(x)$.  
The solutions are differentiable at $x=0$ 
only for discrete binding energies. How are the same eigenvalues selected in the Dirac equation at large $m$?

%
%%%%%%%%%%%%%%%%%%%%%%%%%%%%%%%%%%%%%%%%%%%%%%%%%%%%%%%%
\begin{figure}[h]
\includegraphics[width=\columnwidth]{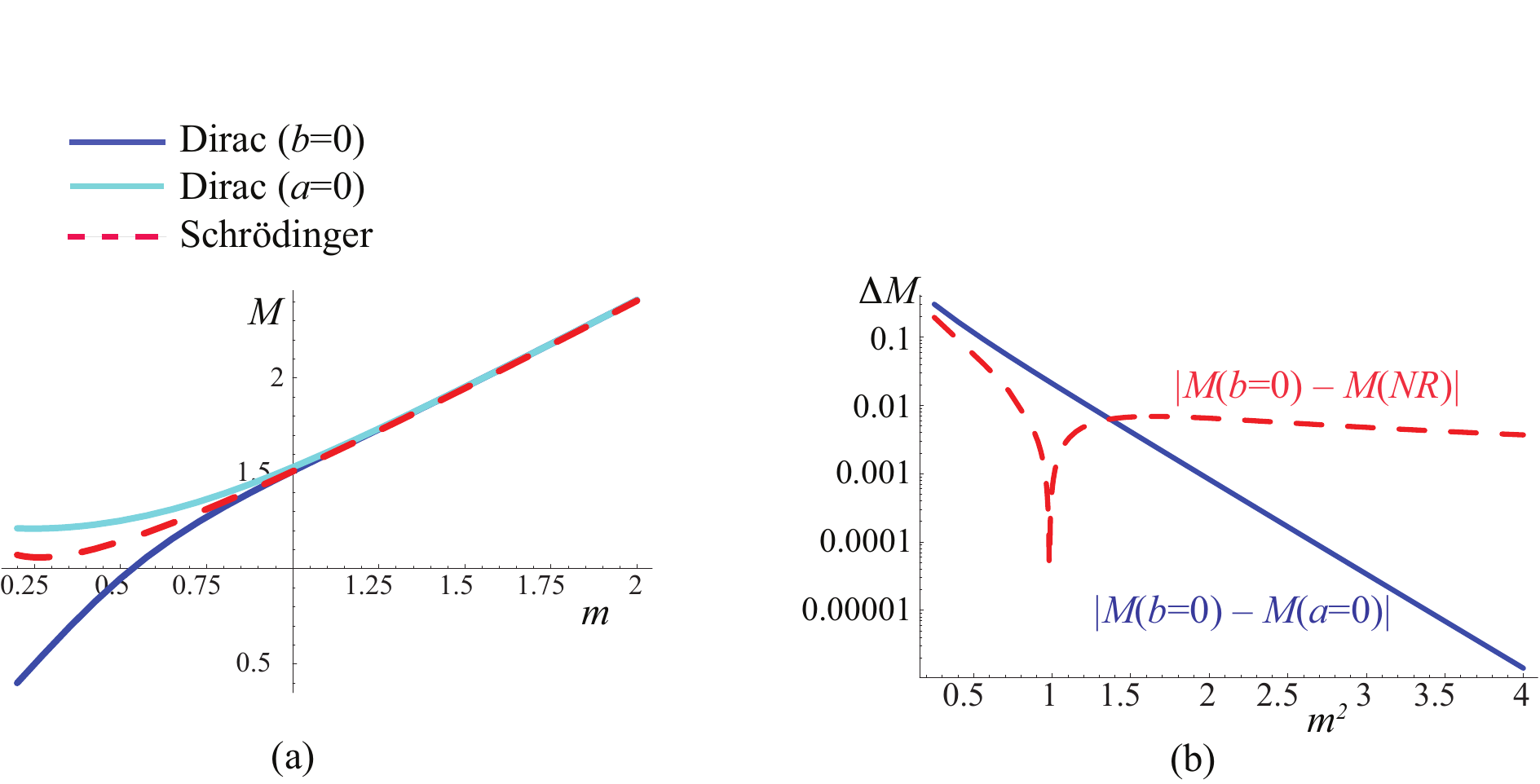}
\caption{Ground state masses $M$ as a function of the constituent fermion mass $m$ (in units of $\sqrt{Ze^2}$). (a) The dark (light) blue solid curves correspond to the Dirac wave function \eq{phisol} with $b=0$ ($a=0$). The dashed red curve shows $M=m+E_b$, with $E_b$ the eigenvalue given by the Schr\"odinger equation \eq{seq} (which is reliable for $m \gsim 1$ only). (b) Absolute values of the ground state mass differences plotted versus $m^2$ on a logarithmic scale. The mass difference between the Dirac solutions for $b=0$ and $a=0$ (solid blue line) decreases exponentially with $m^2$, much faster than the difference of the $b=0$ and Schr\"odinger masses (dashed red curve).{\label{DirMasses}}}
\end{figure}
%%%%%%%%%%%%%%%%%%%%%%%%%%%%%%%%%%%%%%%%%%%%%%%%%%%%%%%%%

It turns out that the two independent solutions in \eq{phisol} become degenerate at large $m$. Setting $M=m+E_b$ in \eq{tdirac} and noting the scaling \eq{nrscaling} we have
\beq
\dsi = \big[m^2+m(2E_b-|x|)\big]+\morder{m^{-2/3}}
\eeq
At large $m$ we may use a stationary-phase approximation in the integral representation of the hypergeometric functions in \eq{phisol}, giving
\beq\label{dirnrlimit}
\psi(\dsi) = (1+i)(a+b)\sqrt{\pi}m^{1/3}e^{\pi m^2/2+i(m^2-\pi/4)}\mathrm{Ai}\big[m^{1/3}(|x|-2E_b)\big]\Big[1+\morder{m^{-2/3}}\Big] ,
\eeq
which agrees with the standard solution \eq{seqAi} of the Schr\"odinger equation. Consequently the nonrelativistic limit of the general solution \eq{phisol} does not depend on the ratio $a/b$.

As seen from \fig{DirMasses} the Dirac eigenvalues are insensitive to $a/b$ already for $m \gsim 1$, where they merge with the bound-state mass given by the Schr\"odinger equation. At large $m$ there is a very narrow range of $a/b$ where the (real or imaginary parts of the) two terms on the right-hand side of \eq{phisol} nearly cancel. Only in this range does the position of the zero, $\dsi=\dsi_0$, depend on the precise value of $a/b$. For, {\it e.g.},
 $m=2.5$ this occurs for the ground state near $a/b=-1.041$, and the width of the interval in $a/b$ that gives a continuum range of masses $M$ is of \order{10^{-6}}. The approximate cancellation in \eq{phisol} then makes the wave function grow very rapidly near $\dsi=\dsi_0$, mimicking the exponential growth of the Schr\"odinger solutions for general $M$. For generic values of $a/b$ and $m \gsim 1$ the bound-state masses given by the Dirac equation agree with those of the normalizable solutions to the Schr\"odinger equation as indicated in \fig{DirMasses}. 

The issue of the approach to nonrelativistic dynamics was also addressed in \cite{titchmarsh}. A measure of the relativistic effects was provided by the distance from the real axis of certain poles related to the Dirac eigenvalues, which was found to be  $\sim \exp(-\kappa m^2)$, with $\pi \leq \kappa \leq 2\pi$. In view of this it is interesting to note that the (typical) difference between the Dirac eigenvalues obtained with different $a/b$ decreases similarly with $m$ (here $\kappa=\pi$), as shown by the solid (blue) line in \fig{DirMasses}(b).

\begin{figure}[h]
\includegraphics[width=10cm]{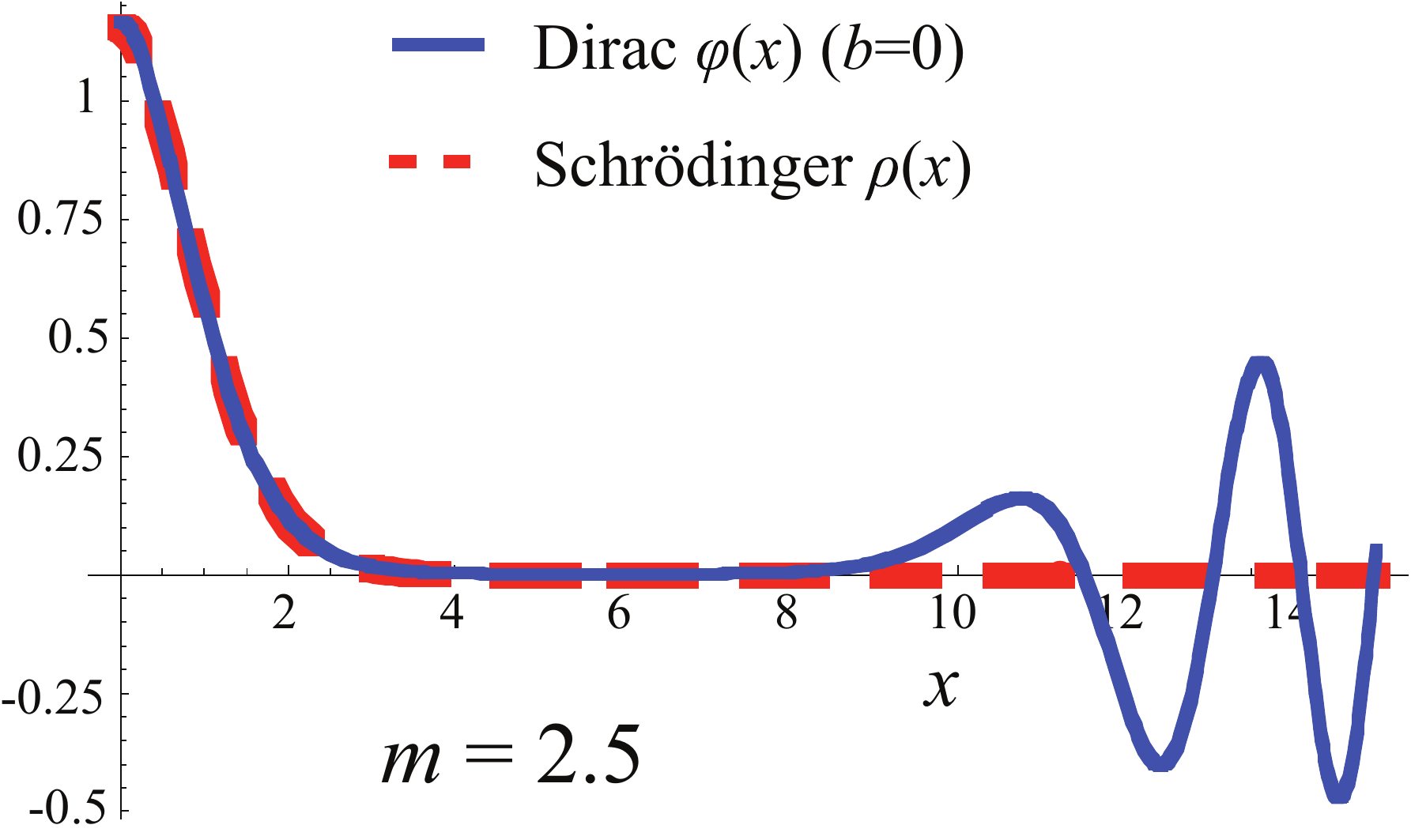}
\caption{\label{Diracfig}The upper component $\vphi(x)$ of the Dirac wave function (continuous blue line) and the Schr\"odinger wave function $\rho(x)$ of \eq{seqAi} (dashed red line). The fermion mass is $m=2.5$ and the parameter $b=0$ in the analytic solution \eq{phisol}. Both solutions are normalized to unity in the range $0 \leq x \leq 6$.}\label{Diracwf}
\end{figure}

As expected, the Dirac wave function is similar to the Schr\"odinger one only for values of $x$ such that $V(x) \ll m$, i.e., for weak binding. Thus the (upper component of) the Dirac wave function agrees with the Schr\"odinger wave function in the nonrelativistic regime, as seen in \fig{Diracwf} (where $m=2.5$ and $b=0$). 
Following its decrease to very small values the Dirac wave function begins to increase at a value of $x$ where 
$V(x) \simeq M$, and becomes \order{1} again, initiating its asymptotic oscillations \eq{asphi} when $V(x) \simeq 2M$.
This is because the wave function depends on $x$ through the variable $\dsi=[M-V(x)]^2$, which takes the same value at $x=0$ as at $V=2M$. The start of the oscillations at a potential energy corresponding to pair production is indicative of the relation between the nonvanishing asymptotic norm and multiparticle effects (the $Z$ 
diagrams mentioned above).

%
%%%%%%%%%%%%%%%%%%%%%%%%%%%%%%%%%%%%%%%%%%%%%%%%%%%%%%%%
%%%%%%%%%%%%%%%%%%%%%%%%%%%%%%%%%%%%%%%%%%%%%%%%%%%%%%%%%

\subsection{Orthogonality  \label{ortho}}
%%%%%%%%%%%%

Two distinct solutions of the Dirac equation \eq{diraccomp}, $\Phi_k=(\vphi_k\ \ \chi_k)^T$ and $\Phi_\ell$ with eigenvalues $M_k \neq M_\ell$ satisfy
\beqa
-i\sigma_1\partial_x\Phi_k+m\sigma_3\Phi_k&=&(M_k-V)\Phi_k, \nn\\
i\partial_x\Phi_\ell^\dag\sigma_1+m\Phi_\ell^\dag\sigma_3 &=&(M_\ell-V)\Phi_\ell^\dag.
\eeqa
Multiplying the first equation by $\Phi_\ell^\dag$ from the left and the second by $\Phi_k$ from the right and subtracting, we find
\beq\label{orthorel1}
-i\partial_x\big(\Phi_\ell^\dag\sigma_1\Phi_k\big)= (M_k-M_\ell)\big(\Phi_\ell^\dag\Phi_k\big).
\eeq
In terms of the solution $\psi(x)=\vphi(x)+\chi(x)$ in \eq{phisol} this is
\beq\label{orthorel2}
\partial_x\left[{\rm Im}(\psi_{\ell}^*\psi_{k})\right] = (M_k-M_\ell) {\rm Re}\left(\psi_{\ell}^*\psi_{k}\right).
\eeq
Recalling that $\vphi(x)=\eta~\vphi(-x)$ is real and $\chi(x)=-\eta~\chi(-x)$ is imaginary both sides of \eq{orthorel2} are odd functions of $x$ if $\eta_k=-\eta_\ell$. Since wave functions of opposite parity are trivially orthogonal it suffices to consider the case $\eta_k=\eta_\ell$. Integrating both sides from $x=0$ to $x=X$ and noting that the insertion at $x=0$ of the left-hand side vanishes we find
\beq\label{orthorel3}
{\rm Im}\big[\psi_{\ell}^*\psi_{k}(x=X)\big] = (M_k-M_\ell) \int_0^X dx {\rm Re}\big[\psi_{\ell}^*\psi_{k}(x)\big].
\eeq
The leading term in the asymptotic limit of \eq{psilimit} is of the form
\beq\label{psilimit2}
\psi(x\to +\infty) = \eta~\psi^*(x\to -\infty) = C\,\dsi^{im^2/2}e^{-i\dsi},
\eeq
where $\dsi=M^2-Mx+x^2/4$, and the complex constant $C$ depends on $m$ as well as on $a$ and $b$, which need not be the same for each bound-state level. For a sufficiently large $X$ that the asymptotic form \eq{psilimit2} of the wave functions applies we have
\beq
\psi_{\ell}^*\psi_{k}(x=X) \simeq C_{k\ell}e^{i(M_k-M_\ell)X},
\eeq
where $C_{k\ell}= C_\ell^* C_ke^{i(M_\ell^2-M_k^2)}$. 
If (to simplify the discussion) 
we choose $X$ such that\footnote{Relaxing this assumption leads to an extra term in Eq.~\eqref{Xintegral}, which is singular at $M_k=M_\ell$ but does not contribute to the final result.}
\beq\label{Xcond}
\cos[(M_k-M_\ell)X]=0
\eeq
\eq{orthorel3} gives, taking into account that the integrand on the right-hand side is symmetric in $x$,
\beq \label{Xintegral}
\int_{-X}^X dx\, {\rm Re}\big[\psi_{\ell}^*\psi_{k}(x)\big] \simeq \frac{2{\rm Re}(C_{k\ell})\sin[(M_k-M_\ell)X]}{M_k-M_\ell}.
\eeq
We recognize the right-hand side as a standard representation of the $\delta$ distribution. Taking $X \to \infty$ we have
\beq\label{ortho4}
\int_{-\infty}^\infty dx\,\Phi_\ell^\dag\Phi_k = \int_{-\infty}^\infty dx\, {\rm Re}\big[\psi_{\ell}^*\psi_{k}(x)\big] = {\rm Re}(C_{k\ell}) 2\pi\delta(M_k-M_\ell).
\eeq

The freedom in choosing the parameter $a/b$ in the general solution \eq{phisol} allows bound-state solutions for a continuous range of masses $M$. Since all these solutions are orthogonal a completeness sum must include them all, as shown in \cite{titchmarsh}. The spectrum of the Dirac equation with a linear potential is thus continuous, similar  
to the case of noninteracting plane waves.

%%%%%%%%%%%%%%%%%%%%%%
\section{Solutions of the two-fermion bound-state equation in $D=1+1$} \label{ffbarsec}
%%%%%%%%%%%%%%%%%%%%%%

In this section we give the general solution for the wave functions $\Psi=e^{i\vphi}\Phi$ that describe the bound states \eq{emustate} of two fermions in $D=1+1$ dimensions,
\beq \label{statedef}
 \ket{E,P}\equiv \int dx_1 dx_2\, \exp\big[\halft iP(x_1+x_2)\big]\bar\psi_1(0,x_1)e^{i\vphi}\Phi(x_1-x_2)\psi_2(0,x_2)\ket{0}_R. 
\eeq
The 2-momentum of the bound state is denoted $(E,P)$, and the Dirac matrices are defined as in \eq{Dmatrices}. The bound-state equation \eq{bse1} is then \cite{Hoyer:2009ep,Dietrich:2012iy}
\beq \label{bse2}
i\partial_x\acom{\sigma_1}{\Phi(x)}-(\partial_x\vphi)\acom{\sigma_1}{\Phi(x)}-\halft P\com{\sigma_1}{\Phi(x)}+m_1\sigma_3\Phi(x)-m_2\Phi(x)\sigma_3 = \big[E-V(x)\big] \Phi(x). 
\eeq
The linear potential imposed by the boundary condition \eq{Fboundcond} is of the form \eq{linpot3d}. We take the coefficient of $\halft |x|$ as our energy scale and thus have the same potential \eq{linpot} as in the Dirac equation, now with $x=x_1-x_2$.

We showed in \cite{Dietrich:2012iy} that $E=\sqrt{P^2+M^2}$ and worked out the $P$ 
 dependence of the wave function. It turned out to be convenient to introduce the variable
\beq\label{trel}
\dsi\equiv(E-V)^2-P^2=M^2-2EV+V^2
 \equiv \dpi^2, 
\eeq
which in the rest frame $(P=0)$ coincides with the variable $\dsi$ defined in \eq{tdirac}, which we used in the Dirac equation\footnote{The invariant mass $M$ is now that of the $f\bar f$ system. The present variable $\dsi$ is related to the variable $s$ of \cite{Dietrich:2012iy} through $\dsi=M^2-2\veps(x)s(x)$.}. 
The expression for $\dsi$ suggests to
define\footnote{Here
the space component of $\dpi$ has the opposite sign compared to its definition in \cite{Dietrich:2012iy}.} the 
``kinematical 2-momentum''  
$\Pi^\mu = P^\mu - eA^\mu$, where $P^\mu = (E,P)^\mu$ is the bound-state momentum and $eA^\mu = (V,0)^\mu$,
\beq\label{kinmom}
\dpi(x) = (E-V(x),P) \equiv (\cosh\zeta,\sinh\zeta) \sqrt{\dsi}.
\eeq
Here the last equality holds when $\dsi= \dpi^2 >0$.
The variable $\dsi(x)=\dsi(-x)$ first decreases with increasing $x>0$,  from $\dsi= M^2$ down to $\dsi=- P^2$ at $V(x)=E$, and then increases with $x$, behaving asymptotically as $\dsi \simeq x^2/4$.

The common phase $\vphi(x)$, which for convenience is extracted from the $2\times 2$ wave function $\Phi$ in the state  \eq{statedef}, is
\beq\label{phasedef}
\vphi(x)= \veps(x)(m_1^2-m_2^2)(\zeta-\xi) = \halft\veps(x)(m_1^2-m_2^2)\log\left[\left(\frac{M}{E+P}\right)^2 \left|\frac{E-V+P}{E-V-P}\right|\right].
\eeq
Here $\zeta$ is defined\footnote{Notice that \eqref{kinmom} defines $\zeta$ only for $\dsi > 0$. The last expression in \eqref{phasedef} can be taken as the definition of $\vphi$ for $\dsi\le 0$, and this is enough to make the wave function $\Phi$ well-defined.} in \eq{kinmom}, and $\xi$ is the rapidity of the bound state,
 $e^{\xi}=(E+P)/M$. 

Taking the complex conjugate of the bound-state equation and changing $x \to -x$ we see that $\Phi(x)$ and $\Phi^*(-x)$ satisfy the same equation. Consequently we may define solutions of definite parity $\eta=\pm 1$ by
\beq\label{Phieta}
\Phi^\eta(x)=\Phi(x)+\eta~\Phi^*(-x),
%\;; 
\hspace{1cm} \Phi^\eta(-x) = \eta~{\Phi^\eta}^*(x).
\eeq
In the following we first construct solutions for $x \geq 0$ and then complete them to the region $x<0$ according to \eq{Phieta}, requiring continuity at $x=0$.

The general structure of a $2\times 2$ wave function $\Phi(x)$ that 
satisfies \eq{bse2} is \cite{Dietrich:2012iy}
\beqa\label{Phidef}
\Phi(x) &=& \phi + \frac{1}{\dsi
}(m_1 \slashed{\dpi}^\dag\, \phi-m_2\phi\, \slashed{\dpi}^\dag), \nn\\
\phi(x) &\equiv& \Phi_0(x)+\Phi_1(x)\sigma_1, 
\eeqa
where $\Phi_0$ and $\Phi_1$ are scalar functions of $x$.
Inserting these expressions into the bound-state equation \eq{bse2} and substituting the variable $\dsi$ of \eq{trel} for $x$ using
\beq \label{txder}
\partial_x =- (E-V)\partial_\dsi \hspace{1cm} (x \geq 0),
\eeq 
we find that $\Phi_0$ and $\Phi_1$ satisfy
\beq\label{bse3}
-2i\partial_\dsi\Phi_1(\dsi) = \left[1-\frac{(m_1-m_2)^2}{\dsi}\right]\Phi_0(\dsi), \hspace{2cm} 
-2i\partial_\dsi\Phi_0(\dsi) = \left[1-\frac{(m_1+m_2)^2}{\dsi}\right]\Phi_1(\dsi). 
\eeq
The explicit dependence on $E$ and $P$ has disappeared, which means that $\Phi_0$ and $\Phi_1$ are the same functions of $\dsi$ in any frame. They are, however, $P$ dependent when viewed as functions of $x$ due to the relation \eq{trel} between $\dsi$ and $x$. The full $2\times2$ wave function $\Phi$ is expressed in terms $\Phi_0$ and $\Phi_1$ by \eq{Phidef}. The relation between the wave function in a frame where the CM momentum is $P$ to the rest frame $(P=0)$ wave function is given by
\beq\label{krel}
\Phi(\dsi) = e^{-\sigma_1\zeta/2}\Phi^{(P=0)}(\dsi) e^{\sigma_1\zeta/2},
\eeq
with $\zeta$ defined by \eq{kinmom} and $\dsi\ge 0$.

\subsection{$f\bar f$ solutions for $m_1=m_2=m$}
%%%%%%%%%%%%%%%%%%%%%%

We first consider the  equal-mass case, $m_1=m_2=m$. Then the phase $\vphi=0$ in \eq{phasedef} and the coupled equations \eq{bse3} reduce to a second-order equation for $\Phi_1(\dsi)$, which has the form of a Coulomb wave equation,
\beq\label{bse4}
4\partial_\dsi^2\Phi_1+\left(1-\frac{4m^2}{\dsi}\right)\Phi_1=0.
\eeq
The solution will oscillate asymptotically, $\Phi_1(\dsi\to \infty) \sim \exp(\pm i \dsi/2)$, analogously to the behavior of the Dirac wave function \eq{asphi}. The general solution for $\Phi_1$ is
\beq\label{Phi1sol}
\Phi_1(\dsi)= \dsi\, e^{-i\dsi/2}\big[a\,\kum(1-im^2,2,i\dsi)+b\, U(1-im^2,2,i\dsi)\big] ,
\eeq
where $a$ and $b$ are constants and $U(\alpha,\beta,z)$ is the confluent hypergeometric function of the second kind.

The behavior of the $U$ function for small argument,
\beq
U(\alpha,2,z \to 0) = \inv{\Gamma(\alpha)}\inv{z}+\morder{\log z},
\eeq
causes the $2\times 2$ wave function $\Phi(x)$ in \eq{Phidef} to be singular at $\dsi=0$ if $b \neq 0$ in \eq{Phi1sol}: Then $\lim_{\dsi\to 0}\Phi_1(\dsi)
 = -ib/\Gamma(1-im^2)$ is nonvanishing, and the singular factor
$1/\dsi$ is uncanceled in \eq{Phidef}.
Such a singularity at $\dsi=0$ prevents even a {\em local} normalizability of the wave function, and causes the orthogonality integrals \eq{orthocond} (Sec.~\ref{orthosec} below) to diverge. 

The $f\bar f$ bound-state equation thus differs significantly from the Dirac equation \eq{diraccomp}, even though both wave functions are oscillatory at large $|x|$. The general solution for the Dirac wave function is regular for finite $x$ and thus locally normalizable, whereas this is true for the $f\bar f$ wave function only provided $b=0$ in \eq{Phi1sol}.

If we express the bound-state mass as $M = 2m+E_b$ then in the limit of large fermion masses ($m\to\infty$) the binding energy $E_b$ and the coordinate $x$ scale as in \eq{nrscaling} (in the rest frame, $P=0$). Substituting
   $\dsi \simeq  4m^2 + 2m(E_b - {\textstyle\frac{1}{2}}|x|)$ 
in the solution \eq{Phi1sol} and using a stationary-phase approximation in the integral representation of the hypergeometric functions they turn into solutions of the nonrelativistic Schr\"odinger equation,
\begin{equation}\begin{split}
\dsi e^{-i\dsi/2} \kum(1-im^2,2,i\dsi) & = 
\left(\frac{2}{m}\right)^{2/3}e^{\pi m^2}
\, \mathrm{Ai}\left[\left(\halft m\right)^{1/3}
(|x|-2E_b)\right],
\\
\dsi e^{-i\dsi/2}\, U(1-im^2,2,i\dsi) &=
-(2 m^2)^{2/3} 
\frac{\pi\, e^{-\pi m^2}}{\Gamma(1-im^2)}
\left\{
\mathrm{Ai}\left[(\halft m)^{1/3}(|x|-2E_b)\right] + i\,\mathrm{Bi}\left[(\halft m)^{1/3}(|x|-2E_b)\right]\right\},
\end{split}\end{equation}
up to \order{m^{-4/3}}  
corrections. The result for the $U$ function involves the nonnormalizable Airy $Bi$ function.

In order to ensure local normalizability\footnote{The requirement of local normalizability was previously used in \cite{Geffen:1977bh}.}, orthogonality of the lowest-order solutions as well as the correct behavior in the nonrelativistic limit we set $b=0$ and thus consider
\begin{equation}\begin{split}\label{Phiexp}
\Phi_1(\dsi)&= 
N\, \dsi\, 
e^{-i\dsi/2}\kum(1-im^2,2,i\dsi) = \frac{N\sinh(\pi m^2)}{\pi m^2}\,
\dsi  
\,e^{-i\dsi/2} \int_0^1 du\, e^{i\dsi u}u^{-i\moe^2}(1-u)^{i\moe^2} = \Phi_1^*(\dsi), 
\\
\Phi_0(\dsi)&= 
-\Phi_1(\dsi)-2iN  
e^{-i\dsi/2}\kum (1-i\moe^2,1,i\dsi) = -\Phi_0^*(\dsi),
\end{split}\end{equation}
where we assumed the normalization constant $N$ to be real. This makes $\Phi_1(\dsi)$ real for all $\dsi$, as may be seen by a $u \to 1-u$ transformation of its integral representation. Correspondingly, $\Phi_0(\dsi)$ is purely imaginary according to \eq{bse3}.

The asymptotic behavior for large $|\dsi|$ is
\begin{equation}\begin{split}\label{aschi}
\Phi_1(\dsi \to \pm\infty) &\simeq
\sqrt{\frac{2}{\pi}}\,\frac{N}{m}
\sqrt{e^{2\pi m^2}-1}\, 
e^{-\pi\moe^2\theta(-\dsi)} \sin\Big[{\frac{\dsi}{2}}-\moe^2\log(|\dsi|)+\arg\Gamma(1+i\moe^2)\Big][1+\morder{\dsi^{-1}}], 
\\
\Phi_0(\dsi \to \pm\infty) &\simeq
-i\sqrt{\frac{2}{\pi}}\,\frac{N}{m}
 \sqrt{e^{2\pi m^2}-1}\,
 e^{-\pi\moe^2\theta(-\dsi)}\cos\Big[{\frac{\dsi}{2}}-\moe^2\log(|\dsi|)+\arg\Gamma(1+i\moe^2)\Big][1+\morder{\dsi^{-1}}],
\end{split}\end{equation} 
where $\theta(-\dsi)=0\ (= 1)$ for $\dsi>0\ (\dsi<0)$.
Due to the oscillatory behavior of the wave functions, the magnitude of $N$ cannot be fixed by a normalization integral. In Sec.~\ref{dualsec} we show that the normalization of highly excited states may be determined using duality between the contributions of bound states and free fermions to current propagators. 

So far we neglected an $\eps(x)$ factor in~\eqref{txder} and thus assumed\footnote{The solutions actually take the form of~\eq{Phiexp} also for $x<0$ when $m_1=m_2$, but the extension to negative $x$ is still nontrivial as the mapping $\dsi=\dsi(x)$ has a kink at $x=0$.} $x \ge 0$.
 When the solutions for $x<0$ are defined according to the parity constraint \eq{Phieta} the choice of phase indicated in \eq{Phiexp} implies that
\beqa\label{cont}
\Phi_1(-x) &=& \eta~\Phi_1(x), \hspace{2cm} \Phi_0(-x) = -\eta~\Phi_0(x),\nn\\[2mm]
\Phi_1(x=0)&=&0\ \ (\eta=-1),  \hspace{.9cm} \Phi_0(x=0)=0\ \ (\eta=+1).
\eeqa
The latter conditions ensure the continuity at $x=0$ of $\Phi_1(x),\,\Phi_0(x)$ and their derivatives. They also determine the discrete bound-state masses $M$ by the positions of the zeros of $\Phi_0(\dsi)$ and $\Phi_1(\dsi)$ for $\eta=\pm 1$, respectively, through $\dsi(x=0) = M^2$ according to \eq{trel}.

\begin{figure}[h]
\includegraphics[width=\columnwidth]{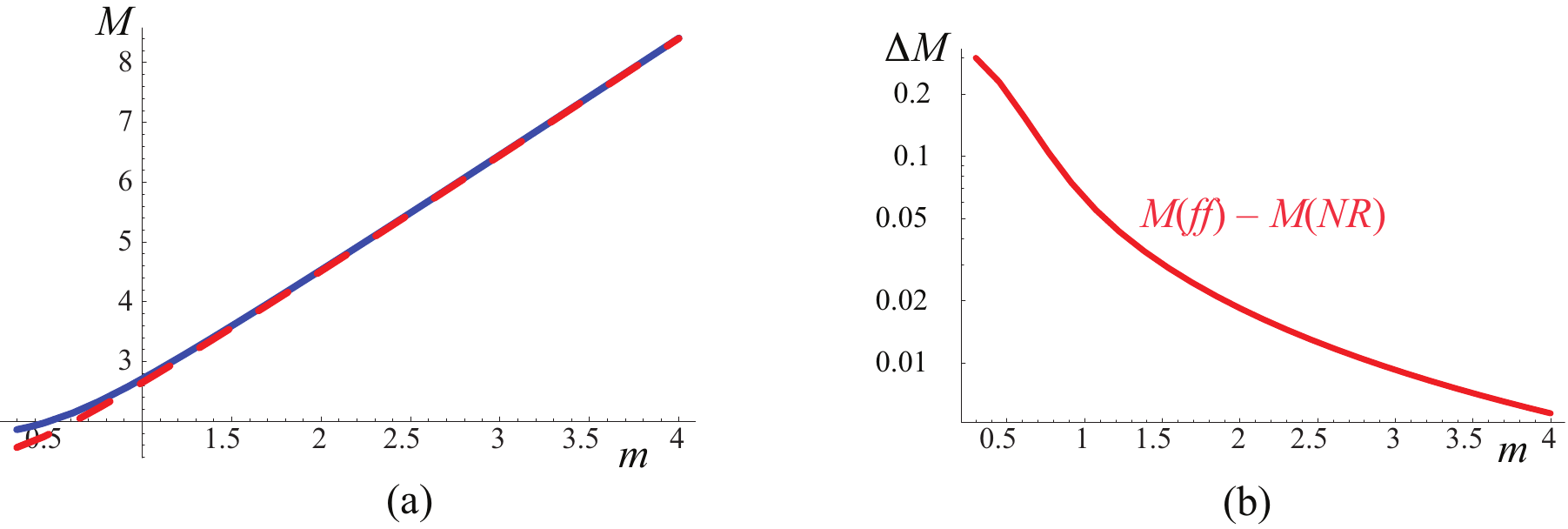}
\caption{The ground state mass $M$ as a function of the fermion mass $m$ [in units of the coefficient of the linear potential \eq{linpot}]. (a) The solid blue curve is for the $f\bar f$ wave function \eq{Phiexp}. The dashed red curve shows $M=2m+E_b$, with $E_b$ the eigenvalue given by the Schr\"odinger equation \eq{seq} (with reduced mass $\halft m$). (b) The difference of the ground state masses in (a), plotted versus $m$ on a logarithmic scale.\label{ffmass}}
\end{figure}

In \fig{ffmass}(a) we compare the symmetric ($\eta=1$) ground state mass as a function of the constituent fermion mass $m$ with the solutions of the Schr\"odinger equation \eq{seqAi} (at the reduced mass $m/2$). As expected there is good agreement for $m \gsim 1$. The mass difference decreases strictly monotonously with $m$, as shown on a logarithmic scale in \fig{ffmass}(b).

%
%%%%%%%%%%%%%%%%%%%%%%%%%%%%%%%%%%%%%%%%%%%%%%%%%%%%%%%%
%%%%%%%%%%%%%%%%%%%%%%%%%%%%%%%%%%%%%%%%%%%%%%%%%%%%%%%%%

%%%%%%%%%%%%%%%%%%%%%%%%%%%%%%%%%%%%%%%%%%%%%%%%%%%%%%%%
\begin{figure}[h]
\includegraphics[width=0.75\columnwidth]{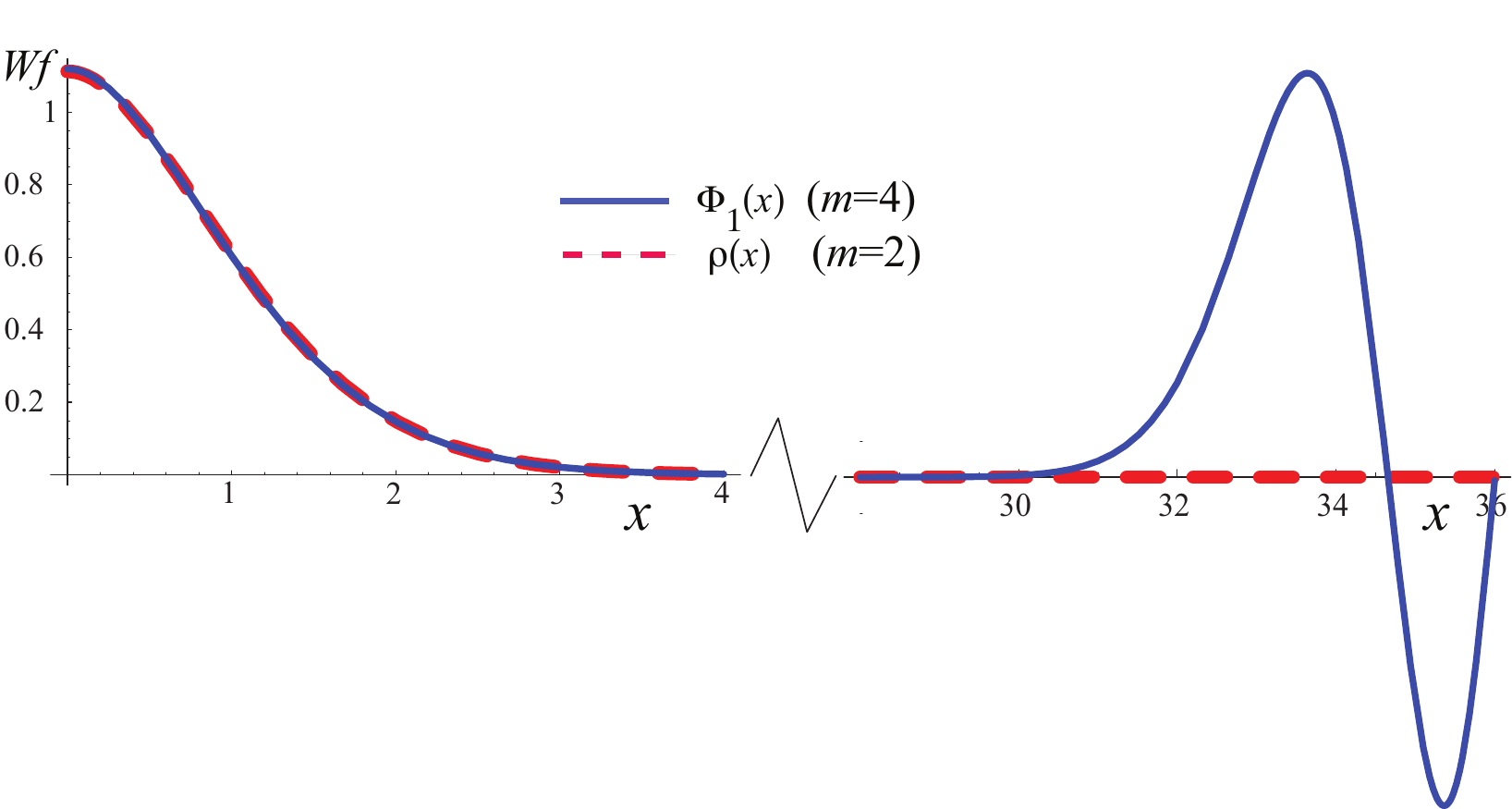}
\caption{The ground state $f\bar f$ wave function $\Phi_1(x)$ in \eq{Phiexp} for $P=0$ and $m=4$ (solid blue line) compared to the nonrelativistic Schr\"odinger wave function $\rho(x)$ in \eq{seqAi} with the reduced mass $m=2$ (dashed red line). The argument of $\Phi_1$ is $\dsi=[M-V(x)]^2$ with $M=8.4100$ while the binding energy for $\rho(x)$ is $E_b=0.4043$. Both wave functions are normalized to unity in the region $0\leq x \leq 5$.}\label{ffwf}
\end{figure}
%%%%%%%%%%%%%%%%%%%%%%%%%%%%%%%%%%%%%%%%%%%%%%%%%%%%%%%%%

In \fig{ffwf} we compare the shape of the $(P=0)$ symmetric $\Phi_1(x)$ ground state wave function in \eq{Phiexp} 
for $m=4$ 
with the corresponding Schr\"odinger wave function \eq{seqAi}.
The two wave functions are nearly equal at low $x$ where $V(x) \ll m$. The relativistic $f\bar f$ wave function increases from small values in the intermediate $x$ region to begin its asymptotic oscillations near $V=2M$. $\Phi_1(x)$ is symmetric in the region $0 \leq V(x) \leq 2M$ since it depends on $x$ only via $\dsi=[M-V(x)]^2$.

\begin{figure}[h]
\includegraphics[width=\columnwidth]{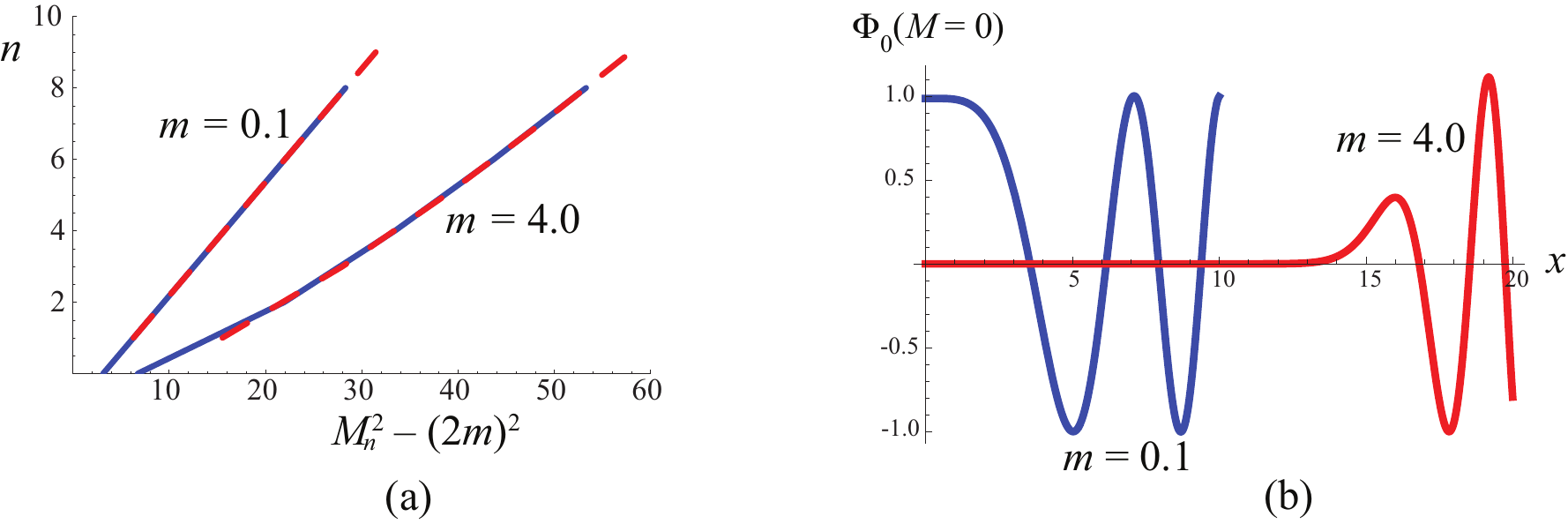}
\caption{(a) Regge trajectories for $f\bar f$ bound states with equal constituent masses, $m=0.1$ (upper curve) and $m=4.0$ (lower curve). The excitation number $n$ of the first five symmetric ($\eta=+1,\ n=0,2,\ldots 8$, blue lines) and antisymmetric ($\eta=-1,\ n=1,3,\ldots 9$, dashed red lines) states are plotted versus $M_n^2-(2m)^2$. (b) The wave function $\Phi_0(x)$ of the $\eta=-1$ state with $M=0$ [not included in (a)] for $m=0.1$ and $m=4.0$. The normalization is chosen arbitrarily such that $\Phi_0=-1$ at the first minimum.}\label{Regge+M0}
\end{figure}

The spectrum of $f\bar f$ states is shown in the form of ``Regge trajectories'' in \fig{Regge+M0}(a), where the square of the bound-state masses $M_n^2$ are shown as a function of the excitation number $n$. The trajectories of the symmetric and antisymmetric $(\eta=\pm 1)$ solutions are degenerate, and in the case of small constituent mass $m=0.1$ almost linear. For $m=4$ the trajectory initially has a smaller slope. 

Notice that the lowest antisymmetric ($\eta=-1$) state has $M=0$ for any $m$, since $\Phi_1(\dsi=0)=0$. This state was not included in \fig{Regge+M0}(a).
\fig{Regge+M0}(b) shows that the wave function $\Phi_0(x)$ of this state is 
essentially nonvanishing only in the relativistic region, $V(x) \gsim 2m$. 
In the nonrelativistic limit the wave function thus tends to zero.

%%%%%%%%%%%%%%%%%%%%%%%%%%%%%%%%%%%%%%%%%%%%%%%%%%%%%%%%
%%%%%%%%%%%%%%%%%%%%%%%%%%%%%%%%%%%%%%%%%%%%%%%%%%%%%%%%%

The mass spectrum can be solved analytically for high excitations, since then $\dsi(x=0) = M^2$ is large at the positions $x=0$, where the wave functions \eq{Phiexp} must vanish according to \eq{cont}, and their asymptotic expressions \eq{aschi} may be used. The zeros are thus approximately at
\beq 
{\frac{\dsi}{2}}-\moe^2\log(\dsi)+\arg\Gamma(1+i\moe^2) = \left\{\begin{array}{cc}  \pi(n+\halft) &  (\eta=+1) \\[2mm] \pi n & (\eta=-1) \end{array}\right..
\eeq
This may be combined into the asymptotic mass spectrum 
\begin{equation}\label{masspect}
 M^2_n = \pi n + 2 m^2  \log (\pi  n)  - 2 \arg \Gamma\left(1+im^2\right) + {\cal O}\left(n^{-1}\right) ,
\end{equation}
where $n$ is odd (even) for $\eta=+1$ ($\eta=-1$).

On the other hand, in the limit of small fermion mass, $m \to 0$, we have $u^{-i\moe^2}(1-u)^{i\moe^2} \simeq 1+i\moe^2\log[(1-u)/u]$ in the integral representation \eq{Phiexp} for $\Phi_1$. Hence we find to \order{\moe^2},
\beq\label{phi1expr}
\Phi_1(\dsi)=N\big(1-\halft\pi \moe^2\big)\left[
2\sin\Big({\frac{\dsi}{2}}\Big)+i\moe^2 \dsi\,
e^{-i\dsi/2}\int_0^1 du\, e^{i\dsi u}\log\Big(\frac{1-u}{u}\Big)+\morder{\moe^4}\right].
\eeq
For $\dsi>0$ the integral can be expressed in terms of sine and cosine integral functions as
\beq
i\dsi 
e^{-i\dsi/2}\int_0^1 du\, e^{i\dsi u}\log\Big(\frac{1-u}{u}\Big) = 2\cos\Big({\frac{\dsi}{2}}\Big)[{\rm Ci}(\dsi)-\log(\dsi)-\gamma_E]+
2\sin\Big({\frac{\dsi}{2}}\Big){\rm Si}(\dsi),
\eeq
where $\gamma_E=0.577216$ is Euler's constant, and
\beq
{\rm Si}(z)=\frac{\pi}{2}-\int_z^\infty du\,\frac{\sin(u)}{u},\hspace{2cm}
{\rm Ci}(z)=-\int_z^\infty du\,\frac{\cos(u)}{u}.
\eeq
Imposing the continuity conditions \eq{cont} at $x=0$ we find the spectrum
\begin{equation}\label{spectrum}
 M^2_n = \pi n + 2 m^2\big[\log(\pi n)-\mathrm{Ci}(\pi n)+ \gamma_E\big] + {\cal O}\left(m^4\right) \ ;  \qquad n = 0,1,2,\ldots, 
\end{equation}
where $n$ is odd (even) for $\eta=+1$ ($\eta=-1$). The case $n=0$ should be understood by taking the limit $n \to 0$, which gives $M_0^2 =0$, the solution shown in \fig{Regge+M0}(b), which is exact for any $m$.

For $m$ exactly equal to zero the full wave function \eq{Phidef} reduces to $\Phi(\dsi)=
-2iN\,\exp(i\sigma_1 \dsi/2)$, 
which is regular at all $\dsi$. Hence there is no constraint on the spectrum when $m=0$. On the other hand, \eq{spectrum} gives $M^2_n = \pi n$ in the $m \to 0$ limit. The discrete spectrum obtained for regular solutions when $m\neq 0$ thus differs, even in the $m \to 0$ limit, from the continuous spectrum found with $m=0$. 
Furthermore, the original bound-state equation \eq{bse2} implied parity doubling when $m_1=m_2=0$: The parity transformed wave function $\gamma^0\Phi(-x)\gamma^0$ is a solution (with $P \to -P$) having the same eigenvalue $E$ as $\Phi(x)$. To the contrary, the $m \to 0$ states have parity $\eta=(-1)^{n+1}$ and are {\em not} parity degenerate.

\subsection{$f\bar f$ solutions for $m_1\neq m_2$}
%%%%%%%%%%%%%%%%%%%%%%

The general solution of \eq{bse3} when
\beq
\dm \equiv m_1^2-m_2^2 \neq 0 
\eeq
is
\beqa\label{genphi}
\Phi_1(\dsi)+\Phi_0(\dsi) &\!=\!& e^{i\dsi/2}\big\{|
\dsi  
|^{-\frac{i}{2}\dm}(a\!+\!ib)m_1\kum(im_2^2,1\!-\!i\dm,-i\dsi)-
|
\dsi  
|^{\frac{i}{2}\dm}(a\!-\!ib)m_2\kum(im_1^2,1\!+\!i\dm,-i\dsi)\big\},\nn\\ \\
\Phi_1(\dsi)-\Phi_0(\dsi)  
&\!=\!& e^{-i\dsi/2}\big\{|
\dsi  
|^{\frac{i}{2}\dm}(a\!-\!ib)m_1\kum(-im_2^2,1\!+\!i\dm,i\dsi)-
|
\dsi  
|^{-\frac{i}{2}\dm}(a\!+\!ib)m_2\kum(-im_1^2,1\!-\!i\dm,i\dsi)\big\},\nn
\eeqa
where $a$ and $b$ are complex constants\footnote{The absolute values in 
$|\dsi|^{\pm i\dm/2}$ 
specify the branch choices for positive and negative $\dsi$.
This choice 
is  
natural in view of our definition of the phase $\vphi$ in
\eqref{phasedef}, which behaves as 
$e^{i\vphi}\sim |\dsi|^{\mp i\dm/2}$ for $E-V \to \pm P$.}.
 The parametrization of the constants was chosen such that for real $a$ and $b$, $\Phi_1$ is real and $\Phi_0$ is imaginary as in the case of equal masses above. We may assume that $x>0$, with the solutions
of definite parity  
defined as in \eq{Phieta}.

The 
$2\times 2$ wave function $\Phi$ is generally singular at $\dsi=0$ according to \eq{Phidef}, which may be expressed as
\beq\label{fullphi}
\Phi=\Phi_0\left\{1+\frac{m_1-m_2}{\dsi}\big[(E-V)\sigma_3+P\,i\sigma_2\big]\right\}+
\Phi_1\left\{\sigma_1+\frac{m_1+m_2}{\dsi}\big[P\,\sigma_3+(E-V)i\sigma_2\big]\right\}.
\eeq 
Noting that $\kum(\alpha,\beta,0)=1$ we find that for $\dsi=[(E-V)+P][(E-V)-P] \to 0$, the most singular terms 
of the full bound-state wave function in \eq{statedef} 
are
\beq\label{tsing}
e^{i\varphi}\Phi(E-V\to\pm P) \sim 
\frac{E-V}{\dsi}\,\dm\, |
\dsi 
|^{\mp i\dm}(a\pm ib)(\sigma_3\pm i\sigma_2).
\eeq 
No choice of the parameters $a,b$ can eliminate the $1/\dsi$ singularity at both $E-V=P$ and $E-V=-P$. The phase $|\dsi|^{\mp i\dm}$, however, ensures the integrability of the wave function when multiplied by any regular function. For example, the orthogonality relations to be discussed below are well-defined.

It is interesting to note that in the infinite momentum frame, $P \to +\infty$ (IMF), the
full wave function 
is regular if $a+ib=0$. This choice removes the $1/\dsi$ singularity in \eq{tsing} at $E-V=P$, while in the IMF $E-V\neq -P$ at finite $x$. 
As $P\to\infty$ both square brackets in \eq{fullphi} approach $E(\sigma_3+i\sigma_2) \equiv E\gamma^+$ and the full $2\times 2$ wave function becomes,
for $b=ia$ in \eq{genphi},
\beq
e^{i\vphi}\Phi_{P\to\infty}(\dsi)=-2ia\,M^{i\dm}\frac{m_1m_2}{1+i\dm}E\gamma^+\,e^{-i\dsi/2}\kum(1-im_2^2,2+i\dm,i\dsi) ,
\eeq
which is indeed regular at $\dsi=0$, and also square integrable in $x$.
While $b=ia$ thus appears to be the most physical choice of parameters, we anyhow continue the discussion assuming generic $a$ and $b$.

By using known identities for the $\kum$ functions it is straightforward to check that in the equal-mass limit $m_1, m_2\to m$ the wave function $\Phi_1(\dsi)$ in \eq{genphi} reduces to
\beq\label{eqmasslimit}
\Phi_1(m_1=m_2=m) = ibm\, \big[  e^{i\dsi/2}\kum(im^2,1,-i\dsi)-e^{-i\dsi/2}\kum(-im^2,1,i\dsi)\big] =
 - b m\, \dsi\,
  e^{-i\dsi/2}\kum(1-im^2,2,i\dsi),
\eeq
which agrees with our previous expression \eq{Phiexp} when $N=-bm$.
The $m_1 \to 0$ limit is also simple since $\kum(\alpha=0,\beta,z)=1$. Thus
\beq\label{m1zero} 
\Phi_j(m_1=0) = {\textstyle\frac12}m_2 \big[-(a-ib)\,|\dsi|^{-\frac{i}{2}m_2^2}e^{i\dsi/2}+(-1)^j\,(a+ib)\,|\dsi|^{\frac{i}{2}m_2^2}e^{-i\dsi/2}\big]\ \ (j=0,1).
\eeq

The definition \eq{Phieta} of $\Phi(x<0)$ requires continuity of $\Phi_0(x=0)$ and $\Phi_1(x=0)$ for the bound-state equation \eq{bse3} to be satisfied at all $x$,%\,;
\beqa
\Phi_j^\eta(x=0) &=& \eta{\Phi_j^\eta}^*(x=0)\label{cont2}\\
&& \hspace{4cm} (j=0,1). \nn \\
\partial_x\Phi_j^\eta(x=0) &=& -\eta{\partial_x\Phi_j^\eta}^*(x=0) \label{cont3}
\eeqa
The condition \eq{cont2} requires that both $\Phi_0(0)$ and $\Phi_1(0)$ are real (imaginary) for $\eta=+1\ (\eta=-1)$. In general, this can be satisfied by adjusting the 
overall phase in \eq{genphi} provided the phase difference of $\Phi_0(0)$ and $\Phi_1(0)$ is 0 or $\pi$. This constraint determines the mass spectrum
for fixed $a/b$. The continuity of the derivatives \eq{cont3} follows from \eq{cont2} and the fact that the wave functions satisfy the bound-state equation \eq{bse3} for $x>0$.

As $m_1 \to m_2$ the phase difference between $\Phi_0$ and $\Phi_1$ approaches $\pm\pi/2$ as seen from Eqs.~\eq{bse3} and \eqref{eqmasslimit}.
Therefore, in the equal-mass case these wave functions cannot have the same phase, instead one of them has to vanish as we found in \eq{cont}. As pointed out above, the same happens if $a$ and $b$ are both real (or, more generally, if they have the same phase, i.e., $a/b$ is real). For $m_1 \approx m_2$ the wave functions have a relative phase of nearly $\pm\pi/2$ everywhere except close to a zero of one of them. Thus the spectrum depends smoothly on $m_1-m_2$. 

%%%%%%%%%%%%%%%%%%%%%%%%%%%%%%%%%%%%%%%%%%%%%%%%%%%%%%%%
\begin{figure}[h]
\includegraphics[width=\columnwidth]{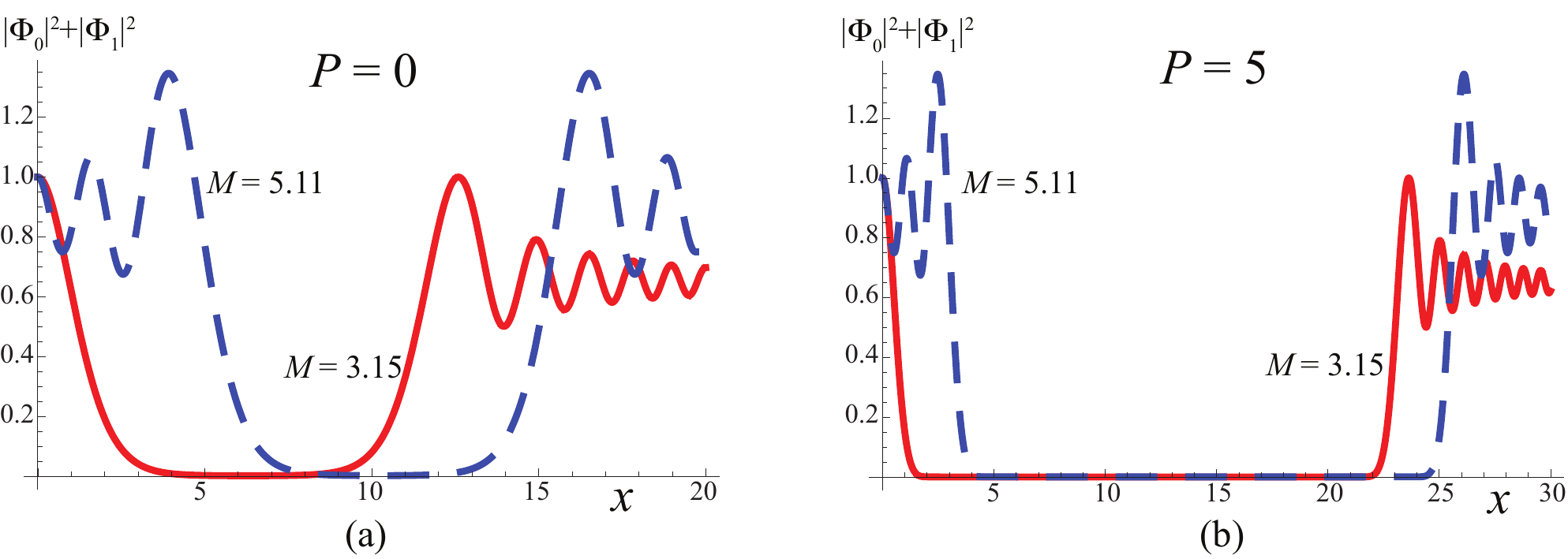}
\caption{(a) The density $|\Phi_0|^2+|\Phi_1|^2$ as a function of the distance $x$ between the constituents for the ground state ($M=3.15$, solid red line) and for an excited state ($M=5.11$, dashed blue line). The constituent masses are $m_1=1.0$ and $m_2=1.5$. (b) The densities in (a) plotted in the case of nonvanishing center-of-mass momentum, $P=5.0$. The densities are symmetric under $x \to -x$ and normalized to unity at $x=0$.}\label{m1m2MP}
\end{figure}
%%%%%%%%%%%%%%%%%%%%%%%%%%%%%%%%%%%%%%%%%%%%%%%%%%%%%%%%%

In \fig{m1m2MP} we illustrate some properties of the solutions in terms of the density $|\Phi_0|^2+|\Phi_1|^2$ for the choice $b=ia$.
 The fermion masses $m_1=1.0,\, m_2=1.5$ are fairly high compared to $V'(x)=0.5$, ensuring that the multipair contributions to the wave functions are well separated in the ground state rest frame [solid red curve in (a)]. The wave function of the excited state [dashed blue curve in (a)] extends to larger fermion separations $x$ before decreasing to small values, but its multipair contributions shift correspondingly in $x$, approximately preserving the extent of the gap. The same densities are shown in (b) for the case of nonvanishing center-of-mass momentum, $P=5$. The wave functions Lorentz contract at low $x$, whereas the length of the gap to the multiparticle contributions grows.

\subsection{Orthogonality of the $f\bar f$ states} \label{orthosec}
%%%%%%%%%%%%%%%%%%%%%%

If we include the phase $\vphi$ in the definition of the $2\times 2$ wave function in the state \eq{statedef},
\beq\label{psidef}
\Psi(x) \equiv e^{i\vphi(x)}\Phi(x),
\eeq
the inner product of two states $k,\,\ell$ reduces to
\beq\label{inprod}
\langle{E_\ell,\,P_\ell}\ket{E_k,\,P_k}= 2\pi\delta(P_k-P_\ell)\int dx\,\tr\big[\Psi_\ell^\dag(x)\Psi_k(x)\big]
\eeq
when only the anticommutators of the fields contribute. 
Thus orthogonality of the states requires
\beq\label{orthocond}
\int dx\,\tr\big[\Psi_\ell^\dag(x)\Psi_k(x)\big]=0 \hspace{1cm} (P_k=P_\ell \equiv P,\ \ E_k\neq E_\ell).
\eeq

The bound-state equations for $\Psi_k(x)$ and $\Psi_\ell^\dag(x)$ are
\beqa\label{bse5}
i\partial_x\acom{\sigma_1}{\Psi_k}-\halft P\com{\sigma_1}{\Psi_k}+m_1\sigma_3\Psi_k-m_2\Psi_k\sigma_3 &=& \big[E_k-V\big] \Psi_k, \nn\\[2mm]
-i\partial_x\acom{\sigma_1}{\Psi_\ell^\dag}+\halft P\com{\sigma_1}{\Psi_\ell^\dag}+m_1\Psi_\ell^\dag\sigma_3-m_2\sigma_3\Psi_\ell^\dag &=& \big[E_\ell-V\big] \Psi_\ell^\dag.
\eeqa
Multiplying the first equation by $\Psi_\ell^\dag$ from the left, the second by $\Psi_k$ from the right and then taking the trace of their difference gives
\beq\label{ortho1}
i\partial_x\tr\Big(\sigma_1\acom{\Psi_\ell^\dag}{\Psi_k}\Big)=(E_k-E_\ell)\tr\big(\Psi_\ell^\dag \Psi_k\big).
\eeq
Integrating both sides over all $x$ the left-hand side gets a contribution only from $x=\pm\infty$ [only the $\Phi_0$ and $\Phi_1$ components of the $2\times 2$ wave function $\Phi$ in \eq{Phidef} contribute on the left-hand side]. To leading order in the $x \to +\infty$ limit we have from \eq{phasedef} and
\eq{genphi}
\beqa
\vphi(x\to\infty) &\simeq& \dm \log[M/(E+P)], \nn\\[2mm]
\Phi_j(\dsi\to\infty) &\simeq&   -\frac12 \big[(-1)^{j}C_1\, 
\dsi 
^{i(m_1^2+m_2^2)/2}e^{-i\dsi/2} + C_2\, 
\dsi
^{-i(m_1^2+m_2^2)/2}e^{i\dsi/2}\big] \hspace{1cm} (j=0,1)\ , \\[2mm]
C_1 &=&  m_1 (a -i b )\frac{ \Gamma\left(1+i\dm\right)}{\Gamma\left(1+i m_1^2\right)}e^{\pi m_2^2/2}
 - m_2 (a +i b )\frac{ \Gamma\left(1-i \dm\right)}{\Gamma\left(1+i m_2^2\right)}e^{\pi m_1^2/2}
 \ , \\[2mm] 
C_2 &=& m_2 (a -i b )\frac{   \Gamma\left(1+i \dm\right)}{\Gamma\left(1-i m_2^2\right)}e^{\pi m_1^2/2}
-m_1 (a +i b )\frac{   \Gamma\left(1-i \dm\right)}{\Gamma\left(1-i m_1^2\right)}e^{\pi m_2^2/2}
\ .
\eeqa
The result in the $x\to -\infty$ limit is given by the complex conjugate of the above, since $\Psi(-x)= \eta\Psi^*(x)$ according to \eq{Phieta}. The product $\Psi_\ell^\dag \Psi_k$ in \eq{ortho1} oscillates asymptotically, and its integral may be defined analogously to that of plane waves, e.g., by adding a factor $e^{-\epsilon |x|}$ with infinitesimal $\epsilon >0$. Then the integral of the left-hand side of \eq{ortho1} vanishes, and the orthogonality \eq{orthocond} is ensured.

%%%%%%%%%%%%%%%%%%%%%%
\section{Duality} \label{dualsec}
%%%%%%%%%%%%%%%%%%%%%%

\subsection{Wave function normalization} \label{normsec}
%%%%%%%%%%%%%%%%%%%%%%

The bound-state equation does not determine the overall normalization (or phase) of the wave functions. Due to contributions from an infinite number of particle pairs the integral of the norm of the Dirac-type wave functions diverges. The relative normalization of high-mass states is needed to determine the parton distributions of the bound states in Sec.~\ref{dis}.
Here we shall use an approximate duality relation to determine their normalization. For simplicity we limit ourselves to the equal-mass case in the following, $m_1=m_2=m$.

%%%%%%%%%%%%%%%%%%%%%%%%%%%%%%%%%%%%%%%%%%%%%%%%%%%%%%%%
\begin{figure}[h]
\includegraphics[width=\columnwidth]{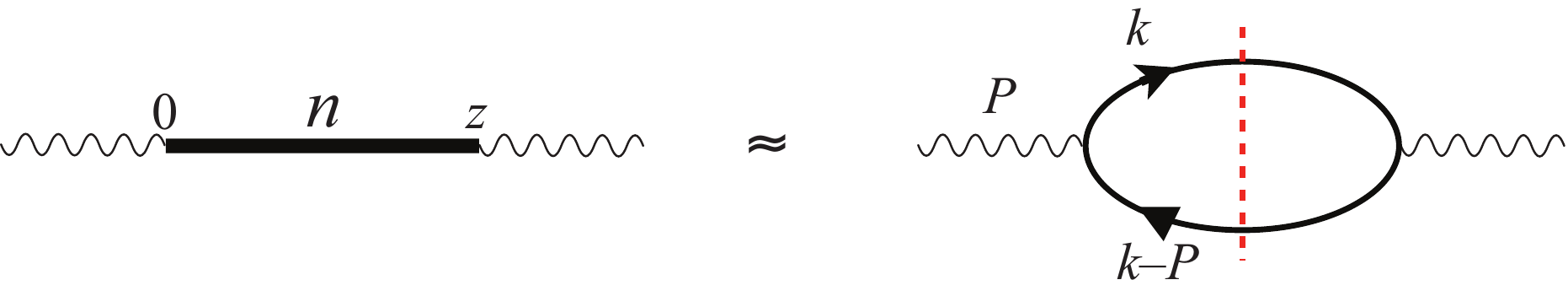}
\caption{ Duality between resonance and fermion-loop contributions to the imaginary part of a current propagator. The relation should hold in a semi-local sense, and become more accurate
at high excitations.
}\label{dual}
\end{figure}
%%%%%%%%%%%%%%%%%%%%%%%%%%%%%%%%%%%%%%%%%%%%%%%%%%%%%%%%%

In our present approximation the bound-state spectrum is
a sequence of zero-width resonances. The (properly averaged) bound-state contribution to the imaginary part of a current propagator is expected to equal the contribution of a free fermion loop, as illustrated in \fig{dual}. This allows to express the wave function at the origin in terms of a calculable perturbative loop contribution. We get consistent normalizations in all frames for scalar, pseudoscalar, vector and pseudovector currents. Furthermore,
in the next subsection we show 
that the bound-state wave functions of
highly excited 
states agree with
those 
of free fermions also for finite separations provided that the potential is much smaller than the energy, $V \ll E$.

Denoting the 2-momentum of the current by $P=(P^0>0,P^1)$ the fermion-loop amplitude in \fig{dual} is for a vector current
\beq\label{vecloop}
L^{\mu\nu}(P) = i\int d^2z \bra{0}\T\big[j^\mu(z)j^\nu(0)\big]\ket{0}e^{iP\cdot z} = i\int\frac{d^2k}{(2\pi)^2} \tr\Big[\frac{\ksl+m}{k^2-m^2+\ieps}\gamma^\mu \frac{\ksl-\Psl+m}{(k-P)^2-m^2+\ieps}\gamma^\nu\Big].
\eeq
The imaginary parts of the loop contribution for this as well as the scalar, pseudoscalar, and pseudovector currents are\footnote{In $D=1+1$ dimensions the pseudovector and vector currents are related, $\bar\psi\gamma_5 \gamma^\mu\psi=\epsilon^{\mu\nu}\bar\psi\gamma_\nu\psi$, where $\gamma_5=\gamma^0\gamma^1$ and $\epsilon^{01}=1$. For completeness we anyhow give results for both.}
\beqa\label{currents}
\im L^{\mu\nu}(P) &=& \Big(-g^{\mu\nu}+\frac{P^\mu P^\nu}{P^2}\Big)\frac{2m^2}{\sqrt{P^2(P^2-4m^2)}} \qquad (\mathrm{vector}:\ j^\mu = \bar\psi\gamma^\mu\psi), \\[2mm]
\im L_S(P)&=& \frac{\sqrt{P^2-4m^2}}{2\sqrt{P^2}} \hspace{4.3cm} (\mathrm{scalar}:\ j_S = \bar\psi\psi), \label{scalar} \\[2mm]
\im L_5(P)&=& \frac{\sqrt{P^2}}{2\sqrt{P^2-4m^2}} \hspace{4.1cm}  (\mathrm{pseudoscalar}:\ j_5 = \bar\psi\gamma_5\psi), \label{pscalar} \\[2mm]
\im L_5^{\mu\nu}(P)&=& \frac{P^\mu P^\nu}{P^2} \frac{2m^2}{\sqrt{P^2(P^2-4m^2)}} \hspace{2.5cm} (\mathrm{pseudovector}:\ j_5^\mu = \bar\psi\gamma_5\gamma^\mu\psi). \label{psvector}
\eeqa

The contribution of a bound state $\ket{n}$, with $\hat P^\mu\ket{n}=P^\mu_n\ket{n}$, to the imaginary part of the vector current with $P^0>0$ is
\beqa
\im L^{\mu\nu}_n(P) &=& \im \; i\int d^2x e^{i(P-P_n)\cdot x} \bra{0}j^\mu(0)\ket{n,t=0}\bra{n,t=0}j^\nu(0)\ket{0}\theta(x^0) \nn\\
&=& 2\pi^2\delta^2(P-P_n)\bra{0}j^\mu(0)\ket{n}\bra{n}j^\nu(0)\ket{0}. \label{rescon}
\eeqa
The expressions \eq{statedef} and \eq{Phidef} for an equal-mass bound state are
\beqa
\ket{n} &=& \int dx_1dx_2 e^{iP_n^1(x_1+x_2)/2}\bar\psi(0,x_1)\Phi(x_1-x_2)\psi(0,x_2)\ket{0}_R, 
 \\[2mm]
\Phi(x)&=& \Phi_0(x)+\Phi_1(x)\gamma_5 \Big[1 - 
\frac{2m}{\dsi}\slashed{\Pi}^\dag\Big],
\eeqa
where $\gamma_5=\gamma^0\gamma^1$. Using $\tr (\gamma_5\gamma^\mu\gamma^\nu)= 2\epsilon^{\mu\nu}$, where $\epsilon^{01}=1$, we get
\beq
{_R \bra{0}}j^\mu(0)\ket{n} =  {_R \bra{0}}\bar\psi(0)\gamma^\mu\psi(0)\ket{n} = \tr\big[\gamma^\mu\gamma^0\Phi(0)\gamma^0\big] = -\frac{4m}{P^2}\Phi_1(0)\epsilon^{\mu\nu}P_\nu.
\eeq
We note that $  {_R \bra{0}}j^\mu(0)\ket{n}P_\mu =0$ as required by gauge invariance. The sum over intermediate states $\sum_n \ket{n}\bra{n}$ includes an integral over the momentum  $P_n^1$ of the single bound state considered in \eq{rescon}. The (local) average over the energy $P_n^0$ of this bound-state contribution should be dual to the loop contribution \eq{currents},
\beq\label{dualcond}
\inv{\Delta M_n^2\,}\int\frac{d^2P_n}{(2\pi)^2}\, \im L^{\mu\nu}_n(P) \simeq  \im L^{\mu\nu}(P).
\eeq
According to \eq{masspect} the bound-state separation $\Delta M_n^2=M_{n+1}^2-M_n^2 \simeq 2\pi$ at large masses [for states with $\Phi_1(0) \neq 0$]. Eq.~\eq{dualcond} then gives
\beq\label{dualvector}
|\Phi_1(x=0)|^2 \simeq \frac{\pi}{2}\sqrt{\frac{M_n^2}{M_n^2-4m^2}}  \qquad (\mathrm{vector\ current}).
\eeq 
Based on the form of the bound-state equation \eq{bse3} we previously noted that $\Phi_1(\dsi)$ is frame independent. Since $\sigma(x=0)=M^2$ is also frame independent this  implies that $\Phi_1(x=0)$ cannot depend on $P_n^1$. The duality relation \eq{dualvector} satisfies this constraint.

We may determine the normalization analogously using currents with other Lorentz structures. Thus
\beqa
{_R \bra{0}}\bar\psi(0)\psi(0)\ket{n} &=& \tr\big[\gamma^0\Phi(0)\gamma^0\big] = 2\Phi_0(0) \hspace{2.5cm} (\mathrm{scalar}), \\[2mm]
{_R \bra{0}}\bar\psi(0)\gamma_5\psi(0)\ket{n} &=& \tr\big[\gamma_5\gamma^0\Phi(0)\gamma^0\big] = -2\Phi_1(0) \hspace{1.9cm} (\mathrm{pseudoscalar}), \\[2mm]
{_R \bra{0}}\bar\psi(0)\gamma_5\gamma^\mu\psi(0)\ket{n} &=& \tr\big[\gamma_5\gamma^\mu\gamma^0\Phi(0)\gamma^0\big] = -\frac{4m}{P^2}\Phi_1(0)P^\mu \hspace{.8cm} (\mathrm{pseudovector}).
\eeqa 
The duality relation for the scalar current gives
\beq\label{dualscalar}
|\Phi_0(0)|^2 \simeq \frac{\pi}{2} \sqrt{\frac{M_n^2-4m^2}{M_n^2}}  \qquad (\mathrm{scalar\ current}).
\eeq
According to \eq{cont} either $\Phi_0(0)=0$ or $\Phi_1(0)=0$ for any given bound state, so \eq{dualscalar} is consistent with and complements the condition \eq{dualvector} for $\Phi_1(0)$. The duality conditions on $\Phi_1(0)$ obtained using the pseudoscalar and pseudovector currents agree with the vector current relation \eq{dualvector}.

For highly excited states we thus find that $|\Phi_0(0)|^2 \simeq \frac{\pi}{2}$ [$|\Phi_1(0)|^2 \simeq \frac{\pi}{2}$] if $\Phi_1(0)=0$ [$\Phi_0(0)=0$]. Using the asymptotic formula~\eqref{aschi} we find that, up to an irrelevant phase,
\beq \label{Nchoice}
 N = \frac{\pi m}{2\sqrt{e^{2\pi m^2}-1}}
\eeq
in~\eqref{Phiexp}.

\subsection{Wave function in momentum space at large $M$} \label{wfduality}
%%%%%%%%%%%%%%%%%%%%%%

It is interesting to consider the duality indicated in \fig{dual} also more differentially: Does the wave function of
highly excited  
bound states resemble the free fermion  distribution given by the imaginary part of the loop, as expected in the parton model? For large masses, $M \gg m$, we may set $m=0$. In the rest frame
($P^1=0$) 
the free fermion momenta are $k=\halft M(1,\pm 1)$. 

The bound state \eq{statedef} is in the rest frame
\beq\label{reststate}
\ket{M,0}=\int\! dx_1dx_2\!\int\!\frac{dk_1 dk_2}{(2\pi)^2 4|k_1||k_2|}\Big[\bar u(k_1)e^{-ik_1x_1}b^\dag_{k_1} +\bar v(k_1)e^{ik_1x_1}d_{k_1}\Big]\Phi(x_1-x_2)\Big[u(k_2)e^{ik_2x_2}b_{k_2} +v(k_2)e^{-ik_2x_2}d^\dag_{k_2}\Big]\!\ket{0}_R.
\eeq
The $m=0$ spinors
\beq\label{spinors1}
u(k) = \frac{|k|\sigma_3-k\,i\sigma_2}{\sqrt{|k|}} \left(\begin{array}{c} 1 \\  0 \end{array}\right),
\hspace{3cm}
v(k) = -\frac{|k|\sigma_3-k\,i\sigma_2}{\sqrt{|k|}} \left(\begin{array}{c} 0 \\  1 \end{array}\right) = \sigma_1 u(k),
\eeq
satisfy
\beqa\label{spinprop}
\bar u(k)u(k)=\bar u(k)\sigma_1u(k)= \bar v(k)v(k)=\bar v(k)\sigma_1v(k) &=& 0, \nn\\
\bar u(k)v(-k)=-\bar v(k)u(-k) &=& -2k, \nn\\
\bar u(k)\sigma_1v(-k)=-\bar v(k)\sigma_1u(-k) &=& 2|k|.
\eeqa
In terms of the Fourier transform of the wave function,
\beq\label{ftwf}
\Phi(k) \equiv \int dx\, \Phi(x)e^{-ikx} =\int dx\, [\Phi_0(x)+\sigma_1\Phi_1(x)]e^{-ikx},
\eeq
the state \eq{reststate} becomes
\beq\label{reststate3}
\ket{M,0}=\int \frac{dk}{2\pi\, 2|k|}\Big\{\big[-\veps(k)\Phi_0(k)+\Phi_1(k)\big]b^\dag_{k}\,d^\dag_{-k}
-\big[\veps(k)\Phi_0(k)+\Phi_1(k)\big]d_{-k}\,b_{k}\Big\}\ket{0}_R, 
\eeq
where $\veps(k)$ is the sign function. It appears to contain both positive ($b^\dag_{k}\,d^\dag_{-k}$) and negative $(d_{-k}\,b_{k})$ energy modes. The wave function of the negative energy modes, however, turns out to vanish.

For $m_1=m_2=0$ and 
$P^1=0$
the bound-state equation \eq{bse3} is
\beq\label{bse0}
i\partial_x \Phi_0(x) = \halft (M-V)\Phi_1(x), \hspace{2cm}
i\partial_x \Phi_1(x) = \halft (M-V)\Phi_0(x).
\eeq
The solutions with a continuous derivative at $x=0$ are
\begin{equation}\begin{split}\label{comb1}
\Phi_1(x)+\Phi_0(x) &= \sqrt{\frac{\pi}{2}}\exp\Big[-\frac{i}{2}Mx+\frac{i}{8} x^2\veps(x)\Big],
\\
\Phi_1(x)-\Phi_0(x) &= \eta \sqrt{\frac{\pi}{2}}\exp\Big[\frac{i}{2}Mx-\frac{i}{8} x^2\veps(x)\Big], 
\end{split}\end{equation}
with $\Phi_j(-x)=(-1)^{j+1} \eta\Phi_j(x)$ as in \eq{cont}. The normalization was determined (up to a phase) by the duality conditions \eq{dualvector} and \eq{dualscalar}.
In momentum space \eq{ftwf} we get
\begin{equation}\begin{split}
\Phi_1(k)+\Phi_0(k) &=
\sqrt{\frac{\pi}{2}}\int dx \exp\big[-i(\halft M+k)x+i x^2\veps(x)/8\big] \simeq \halft (2\pi)^{3/2} \delta(k+\halft M),
  \label{combk1}\\
\Phi_1(k)-\Phi_0(k) &\simeq
\halft\eta (2\pi)^{3/2} \delta(k-\halft M),
\end{split}\end{equation}
where the approximation is valid in the large mass limit, $M \gg 1$. Consequently
\beq
\veps(k)\Phi_0(k)+\Phi_1(k) = 0,
\eeq
and the bound state \eq{reststate3} reduces to
\beq\label{reststate4}
\ket{M,0}= \frac{\sqrt{2\pi}}{2M}\Big(\eta\, b^\dag_{M/2}\,d^\dag_{-M/2} + b^\dag_{-M/2}\,d^\dag_{M/2}\Big)\ket{0}_R, 
\eeq
with a momentum distribution of free fermions, in agreement with the parton model. The approximation made in \eq{combk1} breaks down at large fermion separations where $V(x) \gsim M$ and the effects of confinement set in. Thus the fermions are approximately free only at shorter distances. 

An analogous study of duality may be made in a frame with nonvanishing CM momentum $P$. The parton state corresponding to \eq{reststate4} is then
\beq\label{movingstate4}
\ket{E,P}= \frac{\sqrt{2\pi}}{2M}\Big(\eta\, b^\dag_{E/2+P/2}\,d^\dag_{-E/2+P/2} + b^\dag_{-E/2+P/2}\,d^\dag_{E/2+P/2}\Big)\ket{0}_R.
\eeq

\section{Electromagnetic form factors
} \label{dissec}
%%%%%%%%%%%%%%%%%%%%%%

\subsection{Definition ($m_1 \neq m_2$)}
%%%%%%%%%%%%%%%%%%%%%%

The electromagnetic current is
\beq\label{currentdef}
j^\mu(z)= \sum_{f=1}^2 e_f \bar\psi_f(z)\gamma^\mu\psi_f(z) = e^{i\hat P\cdot z} j^\mu(0) e^{-i\hat P\cdot z},
\eeq
where $e_f$ is the electric charge of flavor $f$ and $\hat P=(\hat P^0,\hat P^1)$ is the
generator of time and space translations. 
We consider the matrix element of $j^\mu(z)$ between bound states of the form \eq{statedef}
\beq \label{ketk}
 \ket{A(P_a)} = \int dx_1 dx_2\, \exp\big[\halft
iP_a^1(x_1+x_2)\big]\bar\psi_1(0,x_1)\Psi_A(x_1-x_2)\psi_2(0,x_2)\ket{0}_R, 
\eeq
where we used the notation \eq{psidef} for the $2\times 2$ wave function $\Psi_A$, whose structure is given by \eq{phasedef} and \eq{Phidef}. Since the bound states are eigenstates of energy and momentum, $\hat P^\mu \ket{A(P_a)} = P_a^\mu \ket{A(P_a)}$, the form factor can be expressed as
\beq\label{formfac}
F^\mu_{AB}(z) = \bra{B(P_b)}j^\mu(z)\ket{A(P_a)}=e^{i(P_b-P_a)\cdot z}\bra{B(P_b)}j^\mu(0)\ket{A(P_a)},
\eeq
where only anticommutators between the fields of the current with those of the states contribute. In effect, the states $\ket{A}$ and $\bra{B}$ replace 
the free $\ket{in}$ and $\bra{out}$ states of standard perturbation theory. Here the asymptotic states are bound by the instantaneous Coulomb potential \eq{linpot3d} arising from the boundary condition \eq{Fboundcond} on $A^0$
and have no \order{e^2} contributions. We expect that a perturbative expansion can be formulated 
as in \eq{smatrix}.
In the following we restrict ourselves to the lowest-order contribution.

Since the potential is confining we consider only neutral bound states, and thus set $e_1=e_2=1$ in \eq{currentdef}. Then the current couples equally to both flavors,
\beqa\label{ff1}
F^\mu_{AB}(z) &=& \sum_{f=1}^2 F^{(f)\mu}_{AB}(z) = e^{i(P_b-P_a)\cdot z}\int dx_1 dx_2 dy_1 dy_2 e^{i(x_1+x_2) P_a^1/2-i(y_1+y_2) P_b^1/2}\nn\\
&\times& {_R} 
\bra{0}\psi_2^\dag(0,y_2)\Psi_B^\dag(y_1-y_2)\gamma^0 \psi_1(0,y_1)\sum_{f=1}^2\bar\psi_f(0,0)\gamma^\mu\psi_f(0,0) \bar\psi_1(0,x_1)\Psi_A(x_1-x_2)\psi_2(0,x_2)\ket{0}_R 
\nn\\[2mm]
&=&  e^{i(P_b-P_a)\cdot z}\int dx\, e^{i(P_b^1-P_a^1)x/2}\,
\Big\{\tr\big[\Psi_B^\dag(x)\gamma^\mu\gamma^0\Psi_A(x)\big]-\eta_a\eta_b\tr\big[\Psi_B(x)\gamma^0\gamma^\mu\Psi_A^\dag(x)\big]\Big\},
\eeqa
where, in the second $(f=2)$ term, we used $\big(\gamma^\mu\big)^T\gamma^0=\gamma^0\gamma^\mu$ (since $\gamma^0=\sigma_3,\, \gamma^1=i\sigma_2$) and $\Psi(-x) = \eta{\Psi^*(x)}$ according to \eq{Phieta}.

\subsection{Gauge invariance ($D=3+1$)} \label{gaugeinv}
%%%%%%%%%%%%%%%%%%%%%%

Gauge invariance of the form factor \eq{formfac} requires that
\beq\label{gaugecond}
G_{AB}^{(f)}(z) \equiv \partial_\mu^z F_{AB}^{(f)\mu}(z) =0 \hspace{1cm} (f=1,2)
\eeq
separately for the current of each flavor $f$. We shall show that \eq{gaugecond} is a consequence of the bound-state equations satisfied by $\Psi_A$ and $\Psi_B$. Since the derivation is essentially independent of the masses and of the number of space-time dimensions we here consider $m_1\neq m_2$ and $D=3+1$. Then
\beqa\label{fgexpr}
F^{(1)\mu}_{AB}(z) &=& e^{i(P_b-P_a)\cdot z}\int d\xv\, e^{i(\Pv_b-\Pv_a)\cdot\xv/2}\,
\tr\big[\Psi_B^\dag(\xv)\gamma^\mu\gamma^0\Psi_A(\xv)\big], \nn\\
G_{AB}^{(1)}(0) &=& i\int d\xv\, e^{i(\Pv_b-\Pv_a)\cdot\xv/2}\,
\tr\big[\Psi_B^\dag(\xv)(\Psl_B-\Psl_A)\gamma^0\Psi_A(\xv)\big].
\eeqa
Due to translation invariance we set $z=0$ in $G_{AB}(z)$ without loss of generality. The bound-state equations \eq{bse1} for
$\Psi_A(\xv)$ and $\Psi_B^\dag(\xv)$ are
\beqa
i\nv_x\cdot\acom{\gamma^0\gv}{\Psi_A}-\halft \Pv_a\cdot\com{\gamma^0\gv}{\Psi_A}+m_1\gamma^0\Psi_A-m_2\Psi_A\gamma^0 = (E_a-V)\Psi_A, \nn\\
-i\nv_x\cdot\acom{\gamma^0\gv}{\Psi_B^\dag}+\halft \Pv_b\cdot\com{\gamma^0\gv}{\Psi_B^\dag}+m_1\Psi_B^\dag\gamma^0-m_2\gamma^0\Psi_B^\dag = (E_b-V)\Psi_B^\dag.
\eeqa
Multiplying the first equation by $-\Psi_B^\dag$ from the left, the second by $\Psi_A$ from the right, and taking the trace of their sum gives
\beq
-i\nv_x\cdot\tr\left(\gamma^0\gv\acom{\Psi_B^\dag}{\Psi_A}\right)+\halft(\Pv_b-\Pv_a)\cdot\tr\left(\gamma^0\gv\com{\Psi_B^\dag}{\Psi_A}\right)=(E_b-E_a)\tr\left(\Psi_B^\dag\Psi_A\right).
\eeq
Using $\big[\Psi_B^\dag,\Psi_A\big]=\big\{\Psi_B^\dag,\Psi_A\big\}-2\Psi_A\Psi_B^\dag$ and multiplying both sides by $\exp[i(\Pv_b-\Pv_a)\cdot\xv/2]$ we find
\beq
-i\nv_x\cdot\Big[e^{i(\Pv_b-\Pv_a)\cdot\xv/2}\,\tr\left(\gamma^0\gv\acom{\Psi_B^\dag}{\Psi_A}\right)\Big]
= e^{i(\Pv_b-\Pv_a)\cdot\xv/2}\,\tr\big[\Psi_B^\dag(\xv)(\Psl_B-\Psl_A)\gamma^0\Psi_A(\xv)\big].
\eeq
Integrating both sides over $\xv$ the right-hand side becomes $-iG_{AB}^{(1)}(0)$ and the left-hand side vanishes (assuming that the integral over the oscillating wave functions is regularized as $|\xv| \to \infty$, similarly as for plane waves). This proves the gauge condition \eq{gaugecond} for $f=1$. For $f=2$ the gauge term corresponding to \eq{fgexpr} is
\beq
G_{AB}^{(2)}(0) = -i\int d\xv\, e^{-i(\Pv_b-\Pv_a)\cdot\xv/2}\,
\tr\big[\gamma^0(\Psl_B-\Psl_A)\Psi_B^\dag(\xv)\Psi_A(\xv)\big],
\eeq
and the proof that it vanishes is analogous to the above.

\subsection{Form factor for $m_1 = m_2$
 ($D=1+1$)} 
%%%%%%%%%%%

The expression \eq{ff1} for the form factor simplifies in the case of equal masses, $m_1=m_2=m$. Since $\vphi=0$ the structure \eq{Phidef} of the wave function $\Psi=\Phi$ becomes
\beq\label{wf2}
\Phi(x) = \Phi_0(x)+\Phi_1(x)\gamma^0\gamma^1+2m\Phi_1(x)\frac{{\slashed\dpi}^\dag}{\dsi} 
\gamma^0\gamma^1,
\eeq
where $\dpi(x)=(P^0-V(x),P^1)$ is the
kinematical  
2-momentum \eq{kinmom}. The traces in \eq{ff1} are now
\beqa \label{fftraces}
\halft\tr\big[\Phi_B^\dag\Phi_A\big] &=& 
\Phi_{0B}^*\Phi_{0A}+\Phi_{1B}^*\Phi_{1A}\Big[1+ \frac{4m^2}
{\dsi_{a}\dsi_{b}} 
\,\tilde \dpi_{a}\cdot \dpi_{b}\Big], \nn \\[2mm]
-\halft\tr\big[\Phi_B^\dag\gamma^1\gamma^0\Phi_A\big] &=& \Phi_{0B}^*\Phi_{1A}+\Phi_{1B}^*\Phi_{0A}+ \Phi_{1B}^*\Phi_{1A}\frac{4m^2}
{\dsi_{a}\dsi_{b}} 
\,\varepsilon_{\mu\nu} \tilde \dpi_{a}^\mu \dpi_{b}^\nu,
\eeqa
where $\tilde \dpi_{a}=(P_a^0-V(x),-P_a^1)$ and $\varepsilon_{01}=-1$.

The constraint \eq{gaugecond} of gauge invariance implies that the form factor in $D=1+1$ can be expressed as
\beq\label{invff} 
F^\mu_{AB}(q)\equiv\int d^2z F^\mu_{AB}(z)e^{-iq\cdot z}= (2\pi)^2 \delta^2(P_b-P_a-q) \varepsilon^{\mu\nu}q_\nu F_{AB}(Q^2),
\eeq
where $Q^2 = -q^2$.
Solving this for $F_{AB}(Q^2)$ with $\mu=0$, using Eq.~\eqref{ff1} for the left-hand side, and inserting the traces of~\eqref{fftraces}, we obtain
\beq
F_{AB}(Q^2) = 
-4i\frac{1-\eta_a\eta_b}{q^1}\int_0^\infty dx\, \sin\Big(\frac{q^1x}{2}\Big)\Big[\Phi_{0B}^*(x)\Phi_{0A}(x)+\Phi_{1B}^*(x)\Phi_{1A}(x)\Big(1+\frac{4m^2}
{\dsi_{a}\dsi_{b}} 
\,\tilde \dpi_{a}\cdot \dpi_{b}\Big)\Big]. \label{ffsym1}
\eeq
The form factor vanishes unless $\eta_a\eta_b=-1$. Therefore the factor in the square brackets could be taken to be antisymmetric in $x$, which allowed us to restrict the integration to  
positive $x$.

\section{DIS and parton distributions} \label{dis}
%%%%%%%%%%%%%%%%%%%%%%

We consider the cross section for $e(k_1)+A(P_a) \to e(k_2)+B(P_b)$ in the limit where $\xbj=Q^2/(2P_a\cdot q)$ is fixed, with $q=k_1-k_2$ and $Q^2=-q^2$. The ``inclusive'' system is thus a discrete bound state $B$. The cross section is proportional to the square of the form factor $F_{AB}(Q^2)$ defined in \eq{invff}, with $M_b \propto Q$. 

\qquad

There are some peculiarities with DIS in $D=1+1$ as compared to $D=3+1$:
\begin{itemize}
\item
The  very concept of a ``cross section'' is related to transverse size. We may nevertheless define a Lorentz-invariant cross section by analogy to the usual case and then compare parton and bound-state cross sections.

\item
The virtual photon has a finite longitudinal (``Ioffe'') coherence length in the target rest frame, $L_I \simeq Q^{-1}\,\nu/Q = 1/(2m_a \xbj)$. In the absence of transverse dimensions DIS photons
can be  
coherent on several partons at leading twist. The fractional momentum of a struck parton is kinematically constrained to be $\xbj$.

\item
In $D=1+1$ the scattering angle can be $\theta=0$ (forward) or $\theta=\pi$ (backward scattering). At the parton level, forward (elastic) scattering implies $Q^2=0$ and thus is irrelevant for the Bjorken (Bj) limit. In backward scattering $\hat s \simeq - \hat t = Q^2$, which is analogous to standard DIS in the Breit (or brick-wall) frame.

\item Since the coupling $e$ has the dimension of energy in $D=1+1$, the parton-level cross section will on dimensional grounds be suppressed
(compared to $D=3+1$)  
by a factor $e^4/Q^4$.

\item
The backward scattering amplitude of elementary spin-$\halft$ fermions is proportional to the fermion masses.
This further suppresses their scattering cross section.

\end{itemize}

\subsection{The parton distribution}
%%%%%%%%%%%%%%%%%%%%%%%%%%%%%%%%%%%%%%%%%%%%%

The kinematics of DIS in $D=1+1$ is discussed in Appendix~\ref{AppDIS}. 
We find that the parton distribution may be expressed as 
\beq \label{partondist}
 f(\xbj) = \frac{1}{8\pi m^2} \inv{\xbj}|Q^2 F_{AB}(Q^2)|^2,
\eeq
where $m$ is the mass of the target parton and the invariant  
form factor $F_{AB}(Q^2)$ is given 
in~\eqref{ffsym1} (for a neutral $f\bar f$ state with $m_1=m_2=m$). 
 The mass of the inclusive system
in the Bj limit is 
\beq \label{Mbscaling}
M_b^2 = Q^2\Big(\inv{\xbj}-1\Big).
\eeq

In order to calculate the leading twist parton distribution for a neutral two-body state
we  
analyze the expression~\eqref{ffsym1} at large $Q^2$ and $M_b$. 
We work in the Breit frame where $q^0=P_b^0-P_a^0=0$ and $q^1 = -Q$ is large.
The basic expectation is that the Fourier phase in \eq{ffsym1} limits the integration to $x \lsim 1/Q$, which is the Lorentz contracted equivalent of a finite (Ioffe) distance in the target rest frame. The integrand is \order{Q^0} in this region, and the measure adds a factor of $1/Q$, so we obtain a contribution at leading order $F_{AB}(Q^2) \sim 1/Q^2$. Leading contributions can, however, arise also for larger (typically $\sim Q^0$) values of $x$, if the oscillations of the wave functions should cancel the Fourier phase such that a stationary phase arises. Such contributions are analyzed in detail in Appendix~\ref{AppQ2} and shown not to affect the leading result.

It is convenient to introduce a rescaled variable 
\beq
 v = \frac{x Q}{2}.
\eeq
In the Bj limit, taking $v = \morder{Q^0}$ and using the expressions~\eqref{diskin} for the momenta, 
the variable $\dsi$ defined in \eq{trel} is of \order{Q^0} for the target, 
\beq \label{taappr}
 \dsi_a = M_a^2 -\frac{Q}{2\xbj}|x|+\inv{4}x^2 \simeq M_a^2-\frac{|v|}{\xbj} \equiv 
 \tau_a , 
\eeq  
while it is of \order{Q^2} for the final state, 
\beq \label{tbappr}
\dsi_b = Q^2\Big(\inv{\xbj}-1\Big)-\frac{Q}{2\xbj}|x|+\inv{4}x^2 \simeq Q^2\Big(\inv{\xbj}-1\Big)-\frac{|v|}{\xbj} \equiv 
\tau_b . 
\eeq 
Thus we may use the asymptotic forms \eq{aschi} for 
$\Phi_B(\dsi_b)$. 

\begin{itemize}
\item $\eta_b=-1$ 

The condition $\partial_v\Phi_{0B}(v=0)=0$
of \eq{cont} 
determines $M_b$ such that 
$\cos[\halft\tau_b-\moe^2\log(\tau_b)  
+\arg\Gamma(1+i\moe^2)]=1$ at $v=0$ (up to an irrelevant phase). The $v$ dependence of the logarithm is of \order{Q^{-2}} and may be ignored. As the state is highly excited, we may use the normalization from~\eqref{Nchoice}. Thus 
\beqa\label{asbpb}
\Phi_{0B}(\tau_b) &\simeq& - \frac{i\sqrt{2\pi}}{2}\cos\Big(\frac{v}{2\xbj}\Big),\nn\\[2mm]
\Phi_{1B}(\tau_b) &\simeq& - \frac{\sqrt{2\pi}}{2}\sin\Big(\frac{|v|}{2\xbj}\Big).
\eeqa
The expression \eq{ffsym1} for the DIS form factor becomes
\beq \label{ffinv3}
Q^2 F_{AB}(\eta_b=-) \simeq  -4i\sqrt{2\pi}(1+\eta_a)\int_{0}^\infty dv\, \sin v\Big[\cos\Big(\frac{v}{2\xbj}\Big) i\Phi_{0A}(\tau_a) - \sin\Big(\frac{v}{2\xbj}\Big)\Phi_{1A}(\tau_a)\Big(1+
\frac{2\moe^2}{\xbj\tau_a}\Big)\Big]. 
\eeq
For large $v$, $\tau_a \to -\infty$ and we may use\footnote{Recall that we already took $Q \to \infty$, thus more precisely, the asymptotic expressions hold for $Q \gg v \gg 1$.}  
the asymptotic expressions \eq{aschi} also for $\Phi_A$: 
\beqa\label{asbpa}
\Phi_{0A}(\tau_a) &\simeq& 
 -i\sqrt{\frac{2}{\pi}}\,\frac{N}{m} 
\sqrt{e^{2\pi m^2}-1}\,e^{-\pi \moe^2}\cos\Big(\mu(v) -\frac{v}{2\xbj}\Big) ,\nn\\
&& \hspace{8cm}  (v\to\infty)\\
\Phi_{1A}(\tau_a) &\simeq& 
 \sqrt{\frac{2}{\pi}}\,\frac{N}{m} 
 \sqrt{e^{2\pi m^2}-1}\,e^{-\pi \moe^2}\sin\Big(\mu(v) -\frac{v}{2\xbj}\Big) ,\nn
\eeqa
where 
\beq\label{mudef}
 \mu(v) =  -\moe^2 \log \frac{v}{\xbj} + \frac{\Moe^2}{2} + \arg \Gamma(1 + i \moe^2) ,
\eeq
and $\Moe = \Moe_a$ is the target mass.
The term in the square brackets in \eqref{ffinv3} behaves as
\beq
 \cos\Big(\frac{v}{2\xbj}\Big) i\Phi_{0A}(\tau_a) - \sin\Big(\frac{v}{2\xbj}\Big)\Phi_{1A}(\tau_a)\Big(1+
 \frac{2\moe^2}{\xbj\tau_a} 
 \Big) \simeq 
 \sqrt{\frac{2}{\pi}}\,\frac{N}{m} 
 \sqrt{e^{2\pi m^2}-1}\,e^{-\pi \moe^2} \cos \mu(v)
\eeq
for $v \to \infty$. Because the integrand of \eq{ffinv3} oscillates asymptotically, the integral does not converge. It can, however, be defined in the standard fashion by adding a ``convergence factor,'' {\it e.g.}, a factor $e^{-\epsilon v}$ in the integrand, and by taking $\epsilon \to 0$ in the end. 

Recall that we assumed that $F_{AB}(Q^2)$ only receives leading contributions from the region $x\sim 1/Q$ in order to derive the result~\eqref{ffinv3}. The fact that the regularization procedure works  
suggests that this assumption was correct. We analyze this in detail in Appendix~\ref{AppQ2}.

\item $\eta_b=1$ 

The condition $\Phi_{0B}(v=0)=0$ 
of \eq{cont} 
determines $M_b$ such that 
$\sin[\halft\tau_b-\moe^2\log(\tau_b)  
+\arg\Gamma(1+i\moe^2)]=1$ at $v=0$. Hence,
\beqa\label{asbmb}
\Phi_{0B}(\tau_b) &\simeq& - \frac{i\sqrt{2\pi}}{2}\sin\Big(\frac{|v|}{2\xbj}\Big),\nn\\[2mm]
\Phi_{1B}(\tau_b) &\simeq& \frac{\sqrt{2\pi}}{2}\cos\Big(\frac{v}{2\xbj}\Big).
\eeqa
The expression \eq{ffsym1} of the DIS form factor becomes
\beq \label{ffinv4}
Q^2 F_{AB}(\eta_b=+) \simeq  -4i\sqrt{2\pi}(1-\eta_a)\int_{0}^\infty dv\, \sin v\Big[\sin\Big(\frac{v}{2\xbj}\Big) i\Phi_{0A}(\tau_a) + \cos\Big(\frac{v}{2\xbj}\Big)\Phi_{1A}(\tau_a)\Big
(1+\frac{2\moe^2}{\xbj\tau_a}\Big)\Big].  
\eeq

\end{itemize}

\subsection{Numerical evaluation of the parton distribution}
%%%%%%%%%%%%%%%%%%%%%%%%%%%%%%%%%%%%%%%%%%%%

Let us insert the explicit results \eq{Phiexp} in the expression \eq{ffinv3} and evaluate the integral numerically. We limit ourselves to the case $\eta_a=-\eta_b=+1$. The case of opposite parities can be analyzed similarly. Since the integral does not converge we need to subtract the divergent term and treat it separately. We write
\beq
 Q^2 F_{AB}(\eta_a=+) = I_1 + I_2 ,
\eeq
where
\beqa
 I_1 &=&  -8i\sqrt{2\pi}\int_{0}^\infty dv\, \sin v\Big[\cos\Big(\frac{v}{2\xbj}\Big) i\Phi_{0A}(\tau_a) - \sin\Big(\frac{v}{2\xbj}\Big)\Phi_{1A}(\tau_a)
 \Big(1+\frac{2\moe^2}{\xbj\tau_a}\Big)  
 \nn\\ && - 
\sqrt{\frac{2}{\pi}}\,\frac{N}{m}  
 \sqrt{e^{2\pi m^2}-1}\, e^{-\pi \moe^2} \cos \mu(v) \Big], \\[1mm]
I_2 &=&  -\frac{16iN}{m} 
\sqrt{e^{2\pi m^2}-1}\, e^{-\pi \moe^2} \int_{0}^\infty dv\, \sin v \cos \mu(v) .
\eeqa 
The divergence now only appears in the second integral, which may be calculated analytically. The standard regularization yields
\beq \label{I2an}  
 I_2 = -16i  
  \frac{N}{m}\sqrt{e^{2\pi m^2}-1}\, e^{-\pi \moe^2} \cosh\left(\frac{\pi \moe^2}{2}\right) \cos\left( \frac{\Moe^2}{2}+\moe^2\log\xbj\right)\left|
\Gamma(1+i\moe^2)\right|. 
\eeq
The first integral $I_1$ can be calculated numerically\footnote{This integral is still not absolutely convergent, which makes the numerical integration rather tricky. We found the best results by using the ExtrapolatingOscillatory method of NIntegrate in Mathematica. One can also introduce a cutoff at large $v$ and use standard algorithms for the numerical integration.}.
The parton distribution can then be found by using the formula \eq{partondist}.

\begin{figure}
\includegraphics[width=0.5\textwidth]{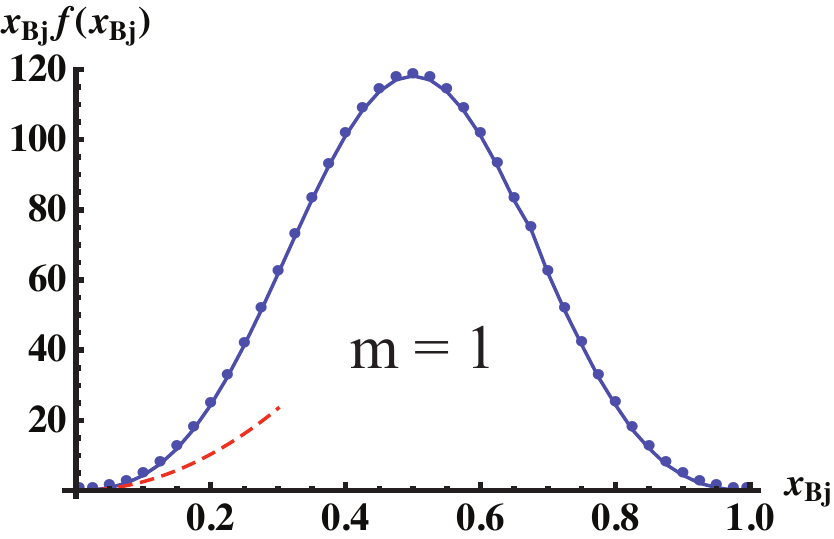}%
\includegraphics[width=0.5\textwidth]{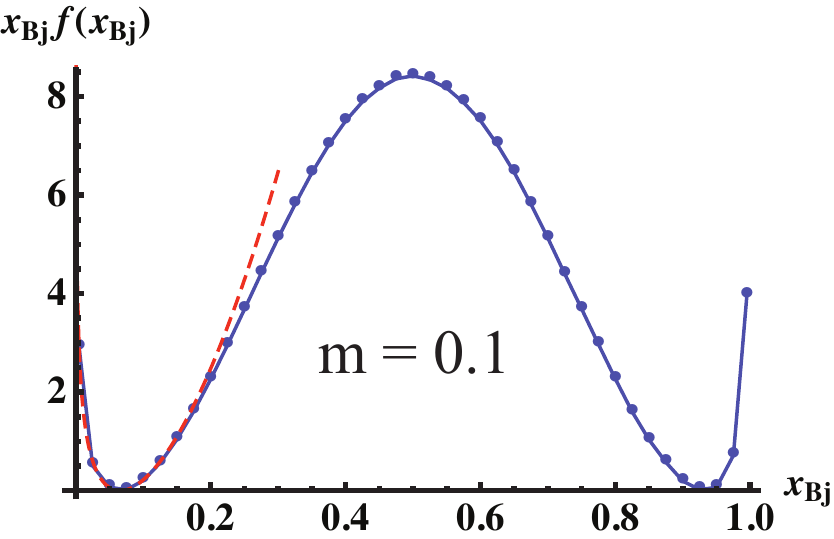}
\includegraphics[width=0.5\textwidth]{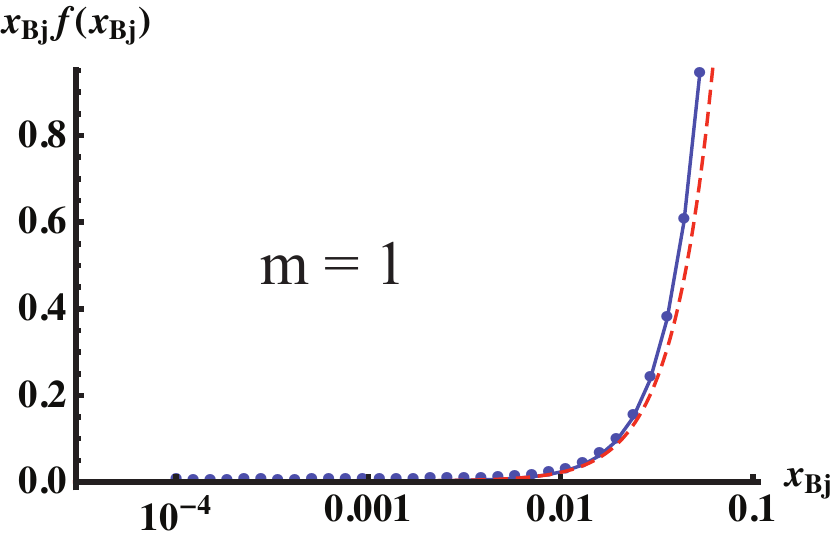}%
\includegraphics[width=0.5\textwidth]{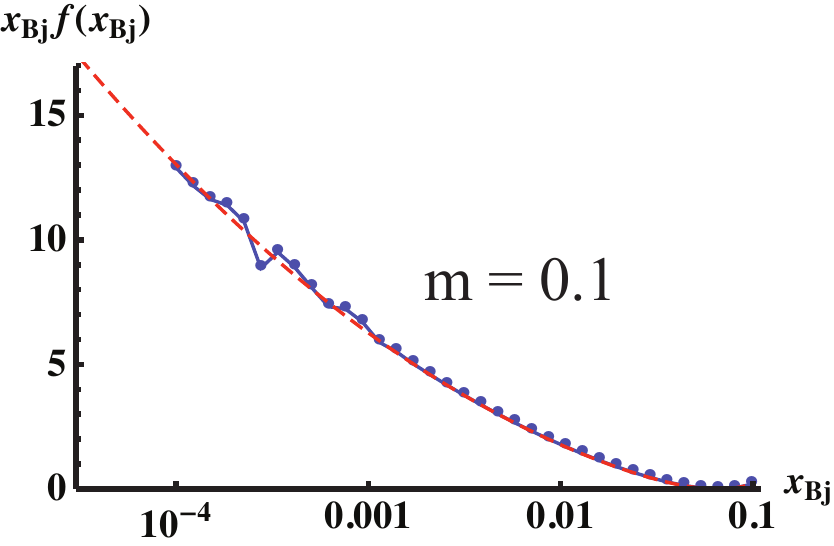}
\caption{\label{xffig}The parton distribution
of the target ground state.  
We plot the result for $\xbj f(\xbj)$ as a function of $\xbj$ in linear scale (top) and in logarithmic scale for low $\xbj$ (bottom). We used $\moe = 1$ (left) and $\moe = 0.1$ (right) for the fermion  
masses. The 
dashed 
red curves show the asymptotic behavior of $f(\xbj)$ up to next-to-leading order as $\xbj \to 0$ [see Eq.~\eq{x0serfin2}].}
\end{figure}

We show numerical results for the parton distribution in Fig.~\ref{xffig}. The target wave function was chosen to be the one with lowest nonzero mass, which indeed has $\eta_a=+1$. We used Eq.~\eqref{Nchoice} for the normalization of the target wave  
function, and chose two reference values $\moe=  0.1$
and $\moe=1.0$ 
for the fermion  
masses. For $\moe=0.1$ the target is highly relativistic, such that $\Moe \simeq 1.78 \gg 2 \moe$, whereas for the second choice $\moe = 1$ we find that $\Moe \simeq 2.70$, {\it i.e.}, a binding energy smaller than the constituent masses. This is reflected in the low-$\xbj$ behavior of $f(\xbj)$, which is qualitatively different in the two cases. The red curves show the
$\xbj \to 0$  
expansion of $f(\xbj)$, which is calculated in Appendix~\ref{AppLowxbj}.

One can check numerically that the integral $I_2$, which arises from asymptotic oscillations of the wave functions, dominates the result for small $\xbj$ and $\moe$. Therefore approximately
\beq
\xbj f(\xbj) \sim \cos^2\left(\Moe^2/2+\moe^2\log\xbj\right)
\eeq
in this region. Inserting the result for the bound-state mass at small $\moe$ with $n=1$ from \eq{spectrum} we find
\beq
 \xbj f(\xbj) \sim \sin^2\left[\moe^2\left(\log\xbj+\log\pi-\mathrm{Ci}(\pi)+\gamma_\mathrm{E}\right)\right] \simeq \moe^4\left(\log\xbj+1.648\right)^2.
\eeq 
Thus the contribution from the oscillations has a node at\footnote{The approximation for the location of the node is poor since terms suppressed by $\xbj$ were neglected.} relatively small $\xbj \simeq 0.19$ when $\moe$ is small, and grows logarithmically $\sim (\log \xbj)^2$ as $\xbj \to 0$, until the distribution ``saturates'' at
$\xbj \sim e^{-1/\moe^2}$.

%%%%%%%%%%%%%%%%%%%%%%%%%%%%%%%%%%%%%%%%%%%%%%%%%%%%%%%%
%%%%%%%%%%%%%%%%%%%%%%%%%%%%%%%%%%%%%%%%%%%%%%%%%%%%%%%%%

%%%%%%%%%%%%%%%%%%%%%%
\section{Concluding remarks} \label{concsec}
%%%%%%%%%%%%%%%%%%%%%%

Analytical studies of bound-state dynamics are usually based on summing Feynman diagrams, e.g., using Dyson-Schwinger techniques \cite{Roberts:2007ji,Alkofer:2009dm}. In the weak coupling limit the dynamics is nonrelativistic and determined by the Schr\"odinger equation. Relativistic, confined states like hadrons may then emerge only when the coupling is strong. In $D=1+1$ dimensions some all-orders, exact results have been obtained, notably for QED at zero fermion mass (the Schwinger model \cite{Schwinger:1962tp}) and for QCD in the limit of a large number of colors, $N_c \to \infty$ (the 't Hooft model \cite{'tHooft:1974hx}). Their study has led to valuable insights -- see, e.g., \cite{Bodwin:1986wq,Einhorn:1976uz,Glozman:2012ev} and references therein. Bosonization 
of the massive Schwinger model furthermore allowed to obtain some approximate results in the strong-coupling limit \cite{Coleman:1976uz}. No analogous results have been found in $D=3+1$ dimensions. Approximations based on a truncation of the 
Dyson-Schwinger equations have allowed analytical studies of hadrons in QCD \cite{Roberts:2007ji,Alkofer:2009dm}, which complement 
first-principles numerical results using lattice methods. Currently holographic approaches to QCD motivated by the AdS/CFT correspondence \cite{Erdmenger:2007cm,deTeramond:2008ht} are under intense study as a means of obtaining results in the strong-coupling limit of the theory.

In this paper we explored a rather different approach to relativistic bound states in gauge field theory. It may be relevant provided the QCD coupling remains perturbative even in the long-distance regime governing hadron binding. Our approach is motivated by features of the data discussed in the Introduction, as well as by phenomenological and theoretical studies which indicate that $\as$ freezes at a moderate value in the infrared \cite{Brodsky:2002nb,Fischer:2006ub,Deur:2008rf,Aguilar:2009nf,Gehrmann:2009eh,Ermolaev:2012xi}. This possibility merits attention since it allows us to bring the powerful techniques of perturbation theory to bear on bound-state dynamics. 

In our scenario the confining potential arises from a boundary condition imposed on the solution of Gauss' law \cite{Hoyer:2009ep}. This gives an exactly linear $A^0$ potential even in $D=3+1$ dimensions, with strength determined by a parameter $\Lambda$ related to the boundary condition. Since $\Lambda$ is independent of $\as$ the potential is of leading order compared to the perturbative \order{\as} interactions. 
We conjecture that perturbative corrections can be systematically included by expanding the time-ordered exponential in the expression \eq{smatrix} of the $S$ matrix. This amounts to developing the perturbative expansion around states bound exclusively by the nonperturbative linear potential. In analogy to the Taylor expansion of ordinary functions, the complete sum formally gives the exact $S$ matrix, independently of the zeroth order configuration.

Previously \cite{Dietrich:2012iy} we verified that the $f\bar f$ Born states are Poincar\'e covariant in $D=1+1$, as expected because the linear potential arises from a boundary condition which is compatible with the field equations of motion. The wave functions turned out to depend on the separation $x$ between the fermions through the frame-invariant square $\dpi^2 =\dsi$ of the ``kinematical momentum'' $\dpi=(E-V,P)$, where $(E,P)$ is the 2-momentum of the bound state and $V(x)$ is the linear potential. An earlier study \cite{Hoyer:1986ei} indicated that Poincar\'e invariance is preserved also in  $D=3+1$.

In this paper we found the analytic expressions of the Born level $f\bar f$ wave functions in $D=1+1$ and studied their properties. Their norm tends to a constant at large $x$. The nearly nonrelativistic case shown in \fig{ffwf} makes it clear that the constant norm reflects fermion pair production at large values of the potential. The norm of the Dirac wave function behaves similarly (\fig{Diracwf}), which in that case implies a continuous Dirac mass spectrum\footnote{The Dirac spectrum is continuous for almost all potentials in both two and four dimensions. Curiously, textbooks rarely mention this fact, even though the exceptional case of the $1/r$ potential in $D=3+1$ is treated in detail.} \cite{nikolsky,sauter,plesset,titchmarsh}. The $f\bar f$ wave function is, however, generally singular at the value of $x$ where the ``kinematical mass'' vanishes, $\dpi^2=0$. Requiring the wave function to be regular gives a discrete spectrum.
A manifestation of the virtual fermion pairs in the bound states is provided by the parton distributions measured by deep inelastic lepton scattering. For relativistic states the parton distribution grows as $x_{Bj} \to 0$, qualitatively in agreement with data on sea quarks.

We found that the wave functions of highly excited bound states agree with parton model expectations in the range of $x$ where the potential is small compared to the bound-state mass: Only fermions and antifermions of positive energy contribute to the bound state, having the momenta of nearly free particles. Duality allows us to determine the overall normalization of the wave functions through the condition that the (average) contribution of the bound states to the imaginary part of current propagators agrees with that of free fermions. The duality relation works in all frames and for all currents.

\acknowledgments
We thank Zbigniew Ambrozi\'nski, Stan Brodsky, and Joachim Reinhardt for discussions. 
Part of this work was done while the authors were visiting or employed by CP$^3$-Origins at the University of Southern Denmark. 
The work of DDD was supported in part by the ExtreMe Matter Institute EMMI and the Helmholtz International Centre (HIC) for FAIR within the framework of the LOEWE program launched by the State of Hesse. 
PH is grateful for the hospitality of CP$^3$-Origins as well as for a travel grant from the Magnus Ehrnrooth Foundation. The work of MJ was in part supported by EU grants PIEF-GA-2011-300984, PERG07-GA-2010-268246 and the EU program ``Thales'' ESF/NSRF 2007-2013.
It was also co-financed by the European Union (European Social Fund, ESF) and
Greek national funds through the Operational Program ``Education and Lifelong Learning'' of the National Strategic
Reference Framework (NSRF) under ``Funding of proposals that have received a positive evaluation in the 3rd and 4th Call of ERC Grant Schemes.''

%%%%%%%%%%%%%%%%%%%%%%%%%%%%%%%%%%%%%%%%%%%%%
%%%%%%%%%%%%%%%%%%%%%%%%%%%%%%%%%%%%%%%%%%%%%
%%%%%%%%%%%%%%%%%%%%%%%%%%%%%%%%%%%%%%%%%%%%%

\appendix
\renewcommand{\theequation}{\thesection.\arabic{equation}}

%%%%%%%%%%%%%%%%%%%%%%%%%%%%%%%%%%%%%%%%%%%%%
\section{DIS in $D=1+1$} \label{AppDIS}
%%%%%%%%%%%%%%%%%%%%%%%%%%%%%%%%%%%%%%%%%%%%%

\subsection{Units and kinematics in $D=1+1$} \label{AppKinematics} 
%%%%%%%%%%%%%%%%%%%%%%%%%%%%%%%%%%%%%%%%%%%%%

The dimension of the Lagrangian is $[\mathcal{L}]=E^2$ in energy units. Hence the fermion and photon fields have $[\psi]=E^{1/2}$ and $[A^\mu]=E^{0}$, while the electron charge\footnote{For clarity we display the charge $e$ explicitly in this appendix, taking $V(x)=\halft e^2 |x|$.}  
$[e] = E^1$. With standard spinor normalizations $\bar u u=-\bar v v = 2m$ the fermion operators have $[b]=[d]=E^0$ and satisfy $\acom{b(k^1)}{b^\dag(p^1)}=2E_p2\pi\delta(k^1-p^1)$. The states have the same dimensions as their creation operators, $[\ket{e^-}]=E^0$. For our bound-state definition \eq{ketk} this implies $[\Phi] = E^1$. 
Consequently the invariant form factor defined in \eq{invff} is dimensionless, $[F_{AB}(Q^2)]=E^0$.

The $2\to 2$ scattering amplitudes
\beq\label{scatamp2}
\mathcal{A}(e\mu\to e\mu)=\bra{e\mu}T\ket{e\mu} = \mathcal{M}(e\mu \to e\mu)(2\pi)^2\delta^2(P_i-P_f)
\eeq
have $[\mathcal{A}]=E^0$ and $[\mathcal{M}]=E^2$. Defining the $2\to 2$ ``cross section'' in analogy to $D=3+1$ as
\beq\label{sigmadef}
\xs(e\mu \to e\mu) = \inv{2s}\int\frac{dp_1^1\,dp_2^1}{(2\pi)^24E_1E_2}|\M|^2 (2\pi)^2 \delta^2(p_1+p_2-P_i)
\eeq
gives $[\xs]=E^0$ as expected. The phase space factor is
\beq\label{sigone}
\xs = \inv{2s}\int\frac{dp_1^1}{4E_1E_2}\delta(E_1+E_2-P^0_i)|\M|^2 =
\inv{2s}\frac{|\M|^2}{4\epsilon_{\mu\nu}p_1^\mu p_2^\nu},
\eeq
where $\epsilon^{01}=1$, and  
we used
\beq
\frac{d(E_1+E_2)}{dp_1^1} = \frac{p_1^1}{E_1}+\frac{p_1^1-P_i^1}{E_2}=\inv{E_1E_2}\big[p_1^1P_i^0-p_1^0 P_i^1\big] = \frac{\epsilon_{\mu\nu}p_1^\mu p_2^\nu}{E_1E_2}.
\eeq
The invariant kinematic factor may be evaluated in the Breit frame where, for $Q \gg m_1,\,m_2$,
\beq
p_1 = \frac{Q}{2}(1,1),\ \ \ p_2 = \frac{Q}{2}(1,-1),
\eeq
giving
\beq\label{sigmaexp}
\xs \simeq \frac{|\M|^2}{4Q^4}\ \ \ {\rm  for}\ Q^2 \to \infty \ \ {\rm  with\ fixed\ } m_1,\, m_2.
\eeq

\qquad

We may evaluate the electron vertex as
\beqa\label{evertex}
V_e^\mu&=& \bar u(k_2)\gamma^\mu u(k_1) = \left(\begin{array}{cc} 1&0 \end{array}\right)
\frac{\ksl_2+m}{\sqrt{E_2+m}}\gamma^\mu \frac{\ksl_1+m}{\sqrt{E_1+m}}
\left(\begin{array}{c} 1 \\  0 \end{array}\right) \nn\\
&=& \inv{\sqrt{(E_1+m)(E_2+m)}}\Big[(E_1+m)k_2^\mu+(E_2+m)k_1^\mu+\halft g^{\mu 0}(k_1-k_2)^2\Big] \equiv \epsilon^{\mu\nu}q_\nu V(Q),
\eeqa
where $q=k_1-k_2$.  
For backward scattering, taking the incoming electron momentum $k_1^1 < 0$ and thus $k_2^1 > 0$, we may evaluate $V_e^1$ explicitly as
\beq
V^1_e = \sqrt{(E_1+m)(E_2-m)}-\sqrt{(E_1-m)(E_2+m)}
=\frac{(E_1+m)(E_2-m)-(E_1-m)(E_2+m)}{\sqrt{(E_1+m)(E_2-m)}+\sqrt{(E_1-m)(E_2+m)}}
=-2m\frac{q^0}{Q} ,
\eeq
which implies
\beq
V(Q) =\frac{2m}{Q}
\eeq
for the invariant part of the vertex in \eq{evertex}. 
Using
\beq
\epsilon^{\mu\rho}q_\rho\,\epsilon^{\nu\sigma}q_\sigma= Q^2 \Big(g^{\mu\nu}-\frac{q^\mu q^\nu}{q^2}\Big)
\eeq
the backward scattering amplitude for $e\mu \to e\mu$ is
\beq\label{fermscatamp}
\M(e\mu \to e\mu) = -e^2V_e(Q)V_\mu(Q)= -4m_em_\mu\frac{e^2}{Q^2}.
\eeq
Using this in the expression \eq{sigmaexp} for the cross section at large $Q^2$ gives
\beq
\xs(e\mu \to e\mu) = 4m_e^2m_\mu^2 \frac{e^4}{Q^8},
\eeq
which is suppressed by $m_e^2m_\mu^2/Q^4$ compared to the ``scaling'' behavior $\sim e^4/Q^4$.

\qquad

We may compare the fermion cross section with that for elementary scalars, which do not have a mass suppression. 
The scalar vertex is
\beq
V_s^\mu = (k_1+k_2)^\mu \equiv \epsilon^{\mu\nu}q_\nu V_s(Q).
\eeq
We can determine $V_s(Q)$ from the space component (taking $k_1^1<0$)
\beq
V_s^1 = \sqrt{E_2^2-m^2}-\sqrt{E_1^2-m^2} = \frac{E_2^2-E_1^2}{k_2^1-k_1^1} = q^0\frac{E_1+E_2}{q^1}=-q^0 V_s(Q)
\eeq
implying
\beq
V_s(Q)= -\frac{E_1+E_2}{k_1^1-k_2^2} = \sqrt{1+\frac{4m^2}{Q^2}}.
\eeq
Consequently the scalar $2 \to 2$ amplitude analogous to \eq{fermscatamp} is of \order{Q^0}, giving a scaling cross section in \eq{sigmaexp}, $\xs \propto e^4/Q^4$.

\subsection{$e(k_1)A(P_a) \to e(k_2) B(P_b)$ in the Bj limit} \label{AppDistribution}
%%%%%%%%%%%%%%%%%%%%%%%%%%%%%%%%%%%%%%%%%%%%%

The Bj limit is defined as usual,
\beq
Q^2 = -(P_b-P_a)^2 \to \infty \ \ \ {\rm with} \ \ \ \xbj = \frac{Q^2}{2P_a \cdot q}\ \ \ {\rm fixed}.
\eeq
The mass $M_a$ of the target $A$ is kept fixed, while the mass of the produced bound state grows with $Q$,
\beq
M_b^2 = (P_a+q)^2 \simeq Q^2\Big(\inv{\xbj}-1 \Big).
\eeq
In the Breit frame,
\beqa\label{diskin}
q = (0,-Q), \hspace{1cm} k_1 &=& \halft Q(1,-1),  \hspace{1cm} k_2=\halft Q(1,1), \nn\\[2mm]
P_a &=& \frac{Q}{2\xbj}(1,1), \hspace{1.0cm} P_b=\frac{Q}{2\xbj}(1,1-2\xbj). 
\eeqa
Using the expression \eq{invff} for the bound-state vertex we find
\beq
\M(eA \to eB) = -e^2\frac{2m_e}{Q}F_{AB}(Q^2).
\eeq
The kinematic factors in the expression \eq{sigone}  
of the cross section are
\beq
s = \frac{Q^2}{\xbj} \ \ \ {\rm and} \ \ \  \epsilon_{\mu\nu}k_2^\mu P_b^\nu = \halft Q^2.
\eeq
Hence the DIS cross section is
\beq
\xs(eA \to eB) = e^4\frac{m_e^2}{Q^6} \xbj |F_{AB}(Q^2)|^2 \equiv 
2\pi e^2 
\frac{d\sigma_{scat}}{dM_b^2},
\eeq
where the second equality implies an average over the bound-state peaks, whose separation in $M_b^2$ is  $2\pi e^2$  
according to \eqref{masspect}.
Converting $dM_b^2 = -Q^2 d\xbj/\xbj^2$ we find the parton distribution $f(\xbj)$ of the target as
\beqa\label{partondistapp}
\frac{d\xs}{d\xbj} &=& \frac{e^2 m_e^2}{2\pi\xbj Q^4}|F_{AB}(Q^2)|^2 \equiv
\hatxs(e\mu \to e\mu) f(\xbj),\nn\\[2mm]
f(\xbj) &=& \frac{1}{8\pi e^2 m^2} \inv{\xbj}|Q^2 F_{AB}(Q^2)|^2,
\eeqa
where $m$ is the mass of the struck fermion in the target $A$.

%%%%%%%%%%%%%%%%%%%%%%%%%%%%%%%%%%%%%%%%%%%%%
\section{Details on the $Q^2 \to \infty$ limit} \label{AppQ2}
%%%%%%%%%%%%%%%%%%%%%%%%%%%%%%%%%%%%%%%%%%%%%

The leading \order{Q^{-2}} result~\eqref{ffinv3},~\eqref{ffinv4} for the DIS form factor $F_{AB}(Q^2)$ in the Bj limit was found by calculating the contribution to the integral~\eqref{ffsym1} for $x \sim 1/Q$ explicitly.
Leading contributions can, in principle, arise also for larger values of $x$, if the oscillations of the wave functions cancel the Fourier phase such that a stationary phase arises. In this appendix we check that such extra contributions are absent.

The form factor  was given in~\eqref{ffsym1} and becomes
\beq
F_{AB}(Q^2) = 
-4i\frac{1-\eta_a\eta_b}{Q}\int_0^\infty dx\, \sin\Big(\frac{Qx}{2}\Big)\Big[\Phi_{0B}^*(x)\Phi_{0A}(x)+\Phi_{1B}^*(x)\Phi_{1A}(x)\Big(1+\frac{4m^2}
{\dsi_a\dsi_b} 
\,\tilde \dpi_{a}\cdot \dpi_{b}\Big)\Big] \label{ffsymapp}
\eeq
in the Breit frame. 
It is necessary to check the behavior of the expression in the square brackets of~\eqref{ffsymapp} by using the asymptotic expressions for $\Phi_j$. 
The wave functions depend on $x$ through the variables $\dsi_{a,b}$. The variable corresponding to the final state is
\beq \label{tbexpr}
 \dsi_b \simeq Q^2\Big(\inv{\xbj}-1\Big)-\frac{Q}{2\xbj}|x|+\inv{4}x^2 ,
\eeq
which  
is large for $x \ge 0$ except  
very close to the roots
\beq \label{xpmdef}
 x_\pm = Q\left[\inv{\xbj}\pm\left(\inv{\xbj}-2\right)\right].
\eeq
The asymptotic  
expansion of $\Phi_B$ can thus be used unless $|x-x_\pm| \lesssim 1/Q$. 
The target wave function depends on
\beq \label{taexpr}
 \dsi_a  = M^2 - \frac{Q}{2 \xbj}|x| + \inv{4}x^2.
\eeq
The asymptotic formulas for $\Phi_A$ can be used unless $x \lesssim 1/Q$ (which was already discussed in the main text) or $|x-2Q/\xbj|\lesssim 1/Q$.

We shall discuss the asymptotics of the wave functions as $\dsi \to -\infty$. To next-to-leading order we find [compare to~\eqref{aschi}]
\beqa
\label{PhiasNLO}
\Phi_1(\dsi) &=& 
\sqrt{\frac{2}{\pi}}\,\frac{N}{m} 
\sqrt{e^{2\pi m^2}-1}\,
e^{-\pi\moe^2} \Bigg\{\sin\big[{\textstyle\frac{\dsi}{2}}-\moe^2\log(-\dsi)+\arg\Gamma(1+i\moe^2)\big] \nn\\
&&+ \frac{m^4 \cos\big[{\textstyle\frac{\dsi}{2}}-m^2 \log(-\dsi)+\arg\Gamma(1+i\moe^2)\big]+m^2 \sin\big[{\textstyle\frac{\dsi}{2}}-\moe^2\log(-\dsi)+\arg\Gamma(1+i\moe^2)\big]}{\dsi}+\morder{\dsi^{-2}} \Bigg\}, \nn\\ 
\nn\\
\Phi_0(\dsi) &=& -i
\sqrt{\frac{2}{\pi}}\,\frac{N}{m} 
\sqrt{e^{2\pi m^2}-1}\,
 e^{-\pi\moe^2} \Bigg\{\cos\big[{\textstyle\frac{\dsi}{2}}-\moe^2\log(-\dsi)+\arg\Gamma(1+i\moe^2)\big]\\\nn
&&- \frac{m^4 \sin\big[{\textstyle\frac{\dsi}{2}}-m^2 \log(-\dsi)+\arg\Gamma(1+i\moe^2)\big]+m^2 \cos\big[{\textstyle\frac{\dsi}{2}}-\moe^2\log(-\dsi)+\arg\Gamma(1+i\moe^2)\big]}{\dsi}+\morder{\dsi^{-2}} \Bigg\}.
\eeqa
Also, the expression appearing in the last term in the square brackets of~\eqref{ffsymapp} can be written as
\beq \label{tfactor}
\frac{4 m^2  }{\dsi_a \dsi_b}\tilde \dpi_a \cdot \dpi_b
 = 2m^2 \frac{\dsi_a+\dsi_b+(1-\xbj)^2 Q^2/\xbj^2}{\dsi_a\dsi_b}.
\eeq
Notice that the apparent singularities of this term as $\dsi_{a,b} \to 0$ are regularized by the zeroes of $\Phi_{1A,B}$ in~\eqref{ffsymapp}.

We start by discussing the contributions from the regions where 
$x \gsim \morder{Q^0}$ and  
the asymptotic formulas~\eqref{PhiasNLO} work. Then the $\dsi_{a,b}$ are \order{Q} or larger. Thus the factor~\eqref{tfactor} is suppressed by $1/Q$, at least. Neglecting this factor, let us first consider the leading terms of the asymptotic expansion~\eqref{PhiasNLO}.
The leading terms in~\eqref{ffsymapp} then combine, through $\sin(\alpha_a)\sin(\alpha_b)+\cos(\alpha_a)\cos(\alpha_b)=\cos(\alpha_a-\alpha_b)$, to give
\beq\label{ffas1}
F_{AB}(Q^2) \sim -i\frac{1-\eta_a\eta_b}{Q}\int dx\, \sin\Big(\frac{Qx}{2}\Big) \cos\Big[{\frac{\dsi_a(x)}{2}}-{\frac{\dsi_b(x)}{2}}+\moe^2\log\frac{\dsi_b(x)}{\dsi_a(x)}\Big]
\eeq
where, for $P_a^0=P_b^0 =E$,
\beq
\dsi_a(x)-\dsi_b(x) = M_a^2-M_b^2
\eeq
is fixed and $x$ independent. The remaining $x$ dependence is logarithmic and cannot cancel the rapidly oscillating Fourier phase\footnote{Notice that this argument does not work when $\dsi_a$ or $\dsi_b$ is close to zero, and the logarithmic term varies rapidly. This happens, however, only in the regions where the asymptotic expansions are not reliable to start with.}. Hence no stationary phase can arise from the leading behavior of the first two terms in the integrand, which suggests 
that
the integral is limited to $xQ$ of \order{1}. 
Notice also that we used the approximation of~\eqref{taappr} and~\eqref{tbappr} in the main text, which is valid when $x \sim 1/Q$. The variation of~\eqref{ffas1} due to this approximation is 
\beq
 \sim \inv{Q} \int dx \, \sin\Big(\frac{Qx}{2}\Big) \frac{x}{Q} \sim \inv{Q^4}
\eeq
and thus indeed subleading.

Let us then discuss the inclusion of the next-to-leading terms in~\eqref{PhiasNLO} or the term proportional to the factor~\eqref{tfactor}. These terms are suppressed by at least one power of $\dsi_a$ or $\dsi_b$, i.e., a power of $1/Q$. As the integral in~\eqref{ffsymapp} is only multiplied by $1/Q$, they could still contribute at \order{Q^{-2}} to the form factor if a suitable stationary phase arises. A stationary phase at a generic $x \sim Q$ is not relevant, because then $\dsi_{a,b} \sim Q^2$, which leads to a suppression of at least $Q^{-3}$. 
As we shall see below, the stationary phases only occur for a range of $x$ with length \order{Q^0}, so additional powers of $Q$ cannot arise from the integration. A  
stationary phase at $x \sim Q^0$ would, however, lead to an extra contribution to the form factor at leading order\footnote{Stationary phases near the roots $x_\pm$ of~\eqref{xpmdef} or near $x=2Q/\xbj$ would also be special.}.

Using the expressions~\eqref{PhiasNLO} and~\eqref{tfactor} to expand the factor in the square brackets of~\eqref{ffsymapp} to next-to-leading order in $1/\dsi$, we observe that all possible combinations of the Fourier phase and the phases of $\Phi_{A,B}$ appear. Neglecting constant and slowly varying factors, they are proportional to $Qx \pm \dsi_a \pm \dsi_b$, where the signs of $\dsi_a$ and $\dsi_b$ can be different. If they are, however, the situation is as in the case of the leading order analysis in~\eqref{ffas1}, and no stationary phases arise. Therefore, we discuss only the phases $Qx\pm(\dsi_a+\dsi_b)$, which were absent at leading order. The locations of the stationary phases are found by solving
\beq
 \frac{d}{dx}\left[Qx\pm(\dsi_a+\dsi_b)\right] = 0.
\eeq
The solutions are given by
\beq
 x \simeq Q \left(\inv{\xbj}\pm 1\right) \equiv \hat x_\pm,
\eeq
i.e., they occur at generic \order{Q} values. Near these points the phases behave as $\sim \sfrac14(x-\hat x_\pm)^2$, such that the stationary phases are limited to regions having lengths $\sim Q^0$, as expected.
We conclude that the next-to-leading terms do not contribute to the form factor at leading order\footnote{Notice that if $1-\xbj \sim Q^{-2}$ so that the final state has a finite mass, one of the stationary phases moves to $x \sim Q^0$ signaling the breakdown of the results~\eqref{ffinv3} and~\eqref{ffinv4}.}.

Finally, let us discuss the contributions to $F_{AB}(Q^2)$ from the regions where the asymptotic formulas do not hold for either of the wave functions $\Phi_{A,B}$ [so that $\dsi_a = \morder{Q^0}$ or 
$\dsi_b = \morder{Q^0}$]. 
The case $x\sim 1/Q$ was discussed in the main text. The other regions are the neighborhoods of $x=x_\pm$ in~\eqref{xpmdef} and $x=2Q/\xbj$, respectively. Let us discuss, for definiteness, $x=2Q/\xbj$. Near this point the asymptotic formulas for $\Phi_B$ are valid, and we can write down an expression similar to~\eqref{ffinv3} and~\eqref{ffinv4},
\beq
 F_{AB}(Q^2) \sim \inv{Q} \int_{|x-2Q/\xbj|\sim 1/Q} dx \sin\Big(\frac{Qx}{2}\Big) \big[\cdots\big],
\eeq
where $\cdots$ stands for a function of $xQ$ which is nontrivial but regular within the region of integration. Shifting the integration variable we obtain
\beq
 F_{AB}(Q^2) \sim \inv{Q} \int_{|x|\sim 1/Q} dx \sin\Big(\frac{Qx}{2}+\frac{Q^2}{\xbj}\Big) \big[\cdots\big] \sim  \inv{Q^2} \int_{|v|\sim 1} dv \sin\Big(v+\frac{Q^2}{\xbj}\Big) \big[\cdots\big].
\eeq
The contribution from this region has thus the leading power behavior $\sim 1/Q^2$, but also involves the large phase factor $Q^2/\xbj$. We interpret that the basically arbitrary phase averages the result to zero. Analogous results are found in the neighborhoods of $x=x_\pm$.

%%%%%%%%%%%%%%%%%%%%%%%%%%%%%%%%%%%%%%%%%%%%%
\section{Asymptotics of the parton distribution at small $\xbj$} \label{AppLowxbj}
%%%%%%%%%%%%%%%%%%%%%%%%%%%%%%%%%%%%%%%%%%%%%

It is possible to calculate the 
$\xbj \to 0$ limit  
of the parton distribution analytically. We again assume that $\eta_a=-\eta_b=+1$  
and start by separating the two contributions in Eq.~\eq{ffinv3}  
as
\beq
 Q^2 F_{AB}(\eta_a=+) = J_1 + J_2, 
\eeq
\beqa
 J_1 &=&  -8i\sqrt{2\pi}\int_{0}^\infty dv\, \sin v\Big[\cos\Big(\frac{v}{2\xbj}\Big) i\Phi_{0A}(\tau_a) - \sin\Big(\frac{v}{2\xbj}\Big)\Phi_{1A}(\tau_a)\Big], \\
\label{J2def}
 J_2 &=&
  \frac{16i\moe^2\sqrt{2\pi}} 
 {\xbj}\int_{0}^\infty dv\, \sin v \sin\Big(\frac{v}{2\xbj}\Big)\frac{\Phi_{1A}(\tau_a)}{\tau_a}, \ 
\eeqa
where $\tau_a$ in \eq{taappr} is a function of $v/\xbj$. It would seem  
that the expansions can be calculated by substituting $v \to \xbj v$ in each integral and then developing the factor $\sin( \xbj v)$ as a series at $\xbj =0$.
This, however, leads to integrals that are divergent for $v \to \infty$.
Instead we expand 
the wave functions up to next-to-leading order for 
$\tau_a \to -\infty$,
\beqa
 \Phi_0(\tau_a) &=& -\frac{i\sqrt{2\pi}}{2}e^{-\pi \moe^2}\cos\Big(\mu(v) -\frac{v}{2\xbj}\Big) \nn\\
&&- \frac{i\moe^2\sqrt{2\pi}}{2}e^{-\pi \moe^2}\Big[\cos\Big(\mu(v) -\frac{v}{2\xbj}\Big)+\Big(\moe^2-\Moe^2\Big)\sin\Big(\mu(v) -\frac{v}{2\xbj}\Big)\Big]\frac{\xbj }{v} + \morder{\frac{\xbj^2}{v^2}} ,\\
 \Phi_1(\tau_a) &=&\frac{\sqrt{2\pi}}{2}e^{-\pi \moe^2}\sin\Big(\mu(v) -\frac{v}{2\xbj}\Big) \nn\\
&&- \frac{\moe^2\sqrt{2\pi}}{2}e^{-\pi \moe^2}\Big[\sin\Big(\mu(v) -\frac{v}{2\xbj}\Big)+\Big(\moe^2-\Moe^2\Big)\cos\Big(\mu(v) -\frac{v}{2\xbj}\Big)\Big]\frac{\xbj }{v} + \morder{\frac{\xbj^2}{v^2}} \ ,
\eeqa
where $\mu(v)$ is given in \eq{mudef}, and we used $N$ from \eqref{Nchoice} for the target wave function.

Let us discuss the integral $J_1$ first. Using the expansions of the wave functions we find
\beq
 J_1 =  -8i\pi e^{-\pi \moe^2}\int_{0}^\infty dv\, \sin v\bigg\{ \cos\mu(v) + \moe^2\Big[\cos\Big(\mu(v) -\frac{v}{\xbj}\Big)+\Big(\moe^2-\Moe^2\Big)\sin\mu(v)\Big]\frac{\xbj}{v}+\morder{\frac{\xbj^2}{v^2}}\bigg\} \ .
\eeq
The integral arising from the first term in the wavy brackets was already evaluated in Eq.~\eq{I2an}. The second term can also be calculated analytically. 
The first term 
in the square brackets contains a rapidly oscillating phase as $\xbj \to 0$.
Hence its leading 
contribution arises from the region $v \sim \xbj$, and is of  
\order{(\xbj)^2}. The  
dominant contributions
to the second term 
have $v \sim (\xbj)^0$, and therefore the result is of \order{\xbj}.  
The \order{\xbj^2/v^2} term  
converges fast enough both as $v \to 0$ and as $v \to \infty$ for us to 
develop 
the sine factor as a series at $v=0$ and  
see that this contribution is of 
\order{(\xbj)^2}. Altogether  
we get
\beqa  
 J_1 &=& -8i\pi e^{-\pi \moe^2} \left|\Gamma(1+i\moe^2)\right| \Bigg[\cosh\left(\frac{\pi \moe^2}{2}\right) \cos\Big( \frac{\Moe^2}{2}+\moe^2\log\xbj\Big) \nn\\
&& \hspace{4cm} + \xbj\Big(\moe^2-\Moe^2\Big)\sinh\left(\frac{\pi \moe^2}{2}\right) \sin\Big( \frac{\Moe^2}{2}+\moe^2\log\xbj\Big) \Bigg]
 +\morder{(\xbj)^2} \ .
\eeqa

The integral $J_2$ can be analyzed similarly. There are, however, complications due to the explicit factor of $1/\xbj$ appearing in the coefficient in \eq{J2def}. First, we need to study the terms of the asymptotic expansion of the integrand up to \order{\xbj^2/v^2}. This is necessary in order to determine all \order{\xbj} contributions to the form factor from the region of asymptotically high $v \sim (\xbj)^0$. Second, also the region with small $v \sim \xbj$, where the wave function $\Phi_{1A}$ cannot be estimated in terms of elementary functions, contributes to the form factor at \order{\xbj}. It is hard to find a closed form expression for this contribution, but we will write it down as an integral below.

Let us start with the contributions arising from the region with $v \sim (\xbj)^0$. Developing the integrand at $v \to \infty$ gives
\beqa \label{J2exp}
J_2 &=& 8i\moe^2\pi e^{-\pi \moe^2}  \int_{0}^\infty dv\, \sin v\Bigg\{ -\Big[\cos\Big(\mu(v)-\frac{v}{\xbj}\Big)-\cos \mu(v) \Big]\frac{1}{v}\\ \nn
&& + (m^2-M^2)\Big[\cos\Big(\mu(v)-\frac{v}{\xbj}\Big)-\cos \mu(v) - m^2\sin\Big(\mu(v)-\frac{v}{\xbj}\Big)+m^2\sin \mu(v) \Big]\frac{\xbj}{v^2}+ \morder{\frac{\xbj^2}{v^3}}\Bigg\} \ .
\eeqa
The integral over the  \order{1/v} and \order{\xbj/v^2} terms can be done analytically\footnote{Notice that the integral over the
\order{\xbj/v^2} 
term is convergent at $v \to 0$ despite the factor of $1/v^2$ thanks to cancellations in the numerator.}.
This contribution is given by
\beqa \label{J2result}
J_{2A} &=& 8ie^2\pi e^{-\pi \moe^2} \left|\Gamma(1+i\moe^2)\right| \bigg\{\sinh\Big(\frac{\pi\moe^2}{2}\Big)\cos\Big( \frac{\Moe^2}{2}+\moe^2\log\xbj\Big) -\xbj e^{-\pi \moe^2/2} \Big[(2\moe^2-\Moe^2)\sin\Big(\frac{\Moe^2}{2}\Big)\nn\\
&&
+\moe^2(\moe^2-\Moe^2)\cos\Big(\frac{\Moe^2}{2}\Big) \Big] + \xbj (m^2-M^2)\cosh\Big(\frac{\pi\moe^2}{2}\Big)\sin\Big( \frac{\Moe^2}{2}+\moe^2\log\xbj\Big)\bigg\} + \morder{(\xbj)^2} .
\eeqa

The remaining contribution comes from small $v \sim \xbj$ \footnote{To be precise, Eq.~\eqref{J2result} already contains some \order{\xbj} contributions from the region $v\sim \xbj$. These are the terms without logarithmic phases.}. It can be isolated by subtracting from the integrand its leading terms given in~\eqref{J2exp}, which leads to
\beqa
J_{2B} &=& \frac{8i\pi\moe^2}{\xbj}\int_{0}^\infty dv\, \sin v \Bigg\{\sqrt{\frac{2}{\pi}}\sin\Big(\frac{v}{2\xbj}\Big)\frac{\Phi_{1A}(\tau_a)}{\tau_a} + e^{-\pi m^2}\Big[\cos\Big(\mu(v)\!-\!\frac{v}{\xbj}\Big)-\cos \mu(v) \Big]\frac{\xbj}{v} \\\nn
&& - (m^2-M^2)e^{-\pi m^2}\Big[\cos\Big(\mu(v)-\frac{v}{\xbj}\Big)-\cos \mu(v) - m^2\sin\Big(\mu(v)-\frac{v}{\xbj}\Big)+m^2\sin \mu(v) \Big]\frac{\xbj^2}{v^2} \Bigg\} .
\eeqa
After scaling the integration variable by $\xbj$ the integral reads
\beqa
J_{2B} &=& 8i\pi\moe^2 \int_{0}^\infty dv\, \sin(\xbj v) \Bigg\{
2\sqrt{\frac{2}{\pi}}\sin\Big(\frac{v}{2}\Big)\frac{\Phi_{1A}(\tau_a=M^2-v)}{M^2-v} 
+e^{-\pi m^2}\Big[\cos\big(\tilde \mu(v)-v\big)-\cos \tilde \mu(v) \Big]\frac{1}{v} \nn\\
&& - (m^2-M^2)e^{-\pi m^2}\Big[\cos\big(\tilde \mu(v)-v\big)-\cos \tilde \mu(v) - m^2\sin\big(\tilde\mu(v)-v\big)+m^2\sin \tilde \mu(v) \Big]\frac{1}{v^2} \Bigg\},
\eeqa
where $\tilde \mu(v) = \mu(\xbj v) = M^2/2 -m^2 \log v$, and therefore the whole expression in the curly brackets 
is independent of $\xbj$. The leading term as $\xbj \to 0$ is then found by expanding the factor $\sin(\xbj v)$ at small $\xbj$ and $v$,
\beq
J_{2B} = iC(m,M)\xbj + \morder{(\xbj)^2} ,
\eeq
where
\beqa \label{Cdef}
C(m,M) &=& 8\pi\moe^2 \int_{0}^\infty dv\, \Bigg\{
2\sqrt{\frac{2}{\pi}}v\sin\Big(\frac{v}{2}\Big)\frac{\Phi_{1A}(M^2-v)}{M^2-v}
+e^{-\pi m^2}\Big[\cos\big(\tilde \mu(v)-v\big)-\cos \tilde \mu(v) \Big] \nn\\
&& - (m^2-M^2)e^{-\pi m^2}\Big[\cos\big(\tilde \mu(v)-v\big)-\cos \tilde \mu(v) - m^2\sin\big(\tilde\mu(v)-v\big)+m^2\sin \tilde \mu(v) \Big]\frac{1}{v} \Bigg\} .
\eeqa

Collecting the results,
\beqa \label{x0serfin}
Q^2 F_{AB}(\eta_a\!=\!+)\! &=& \! J_1+J_{2A}+J_{2B} \nn \\  
&=& \!-8i\pi e^{-3\pi \moe^2/2} \left|\Gamma(1+i\moe^2)\right| \bigg\{ \cos\Big( \frac{\Moe^2}{2}+\moe^2\log\xbj\Big) - \xbj\big(\moe^2-\Moe^2\big) \sin\Big( \frac{\Moe^2}{2}+\moe^2\log\xbj\Big) \nn\\
&& \!\!+ \xbj \Big[(2\moe^2\!-\!\,\Moe^2)\sin\Big(\frac{\Moe^2}{2}\Big)
+\moe^2(\moe^2\!-\!\,\Moe^2)\cos\Big(\frac{\Moe^2}{2}\Big) \Big]\!\bigg\} + i  C(\moe,\Moe) \xbj + \morder{(\xbj)^2} . \label{x0serfin2}
\eeqa
The $\xbj \to 0$ expansion for $\xbj f(\xbj)$ obtained by using this formula in \eq{partondist} is shown by the 
dashed red curves in Fig.~\ref{xffig}. Notice that the contribution from the asymptotic oscillations is suppressed by the factor $e^{-3\pi \moe^2/2}$ in the nonrelativistic limit $m \to \infty$. Such 
a suppression 
factor is absent in $C(m,M)$ of \eqref{Cdef}, which mostly arises from small $v \sim \xbj$, {\it i.e}, the region where the wave function $\Phi_{1A}$ is nonvanishing in the nonrelativistic limit.

\end{document}